\documentclass[fleqn,usenatbib]{mnras}

\usepackage{newtxtext,newtxmath}

\usepackage[T1]{fontenc}

\DeclareRobustCommand{\VAN}[3]{#2}
\let\VANthebibliography\thebibliography
\def\thebibliography{\DeclareRobustCommand{\VAN}[3]{##3}\VANthebibliography}

\usepackage{graphicx}	
\usepackage{amsmath}	

\usepackage{scalerel}
\usepackage{tikz}
\usetikzlibrary{svg.path}

\definecolor{orcidlogocol}{HTML}{A6CE39}
\tikzset{
  orcidlogo/.pic={
    \fill[orcidlogocol] svg{M256,128c0,70.7-57.3,128-128,128C57.3,256,0,198.7,0,128C0,57.3,57.3,0,128,0C198.7,0,256,57.3,256,128z};
    \fill[white] svg{M86.3,186.2H70.9V79.1h15.4v48.4V186.2z}
                 svg{M108.9,79.1h41.6c39.6,0,57,28.3,57,53.6c0,27.5-21.5,53.6-56.8,53.6h-41.8V79.1z M124.3,172.4h24.5c34.9,0,42.9-26.5,42.9-39.7c0-21.5-13.7-39.7-43.7-39.7h-23.7V172.4z}
                 svg{M88.7,56.8c0,5.5-4.5,10.1-10.1,10.1c-5.6,0-10.1-4.6-10.1-10.1c0-5.6,4.5-10.1,10.1-10.1C84.2,46.7,88.7,51.3,88.7,56.8z};
  }
}

\newcommand\orcidicon[1]{\href{https://orcid.org/#1}{\mbox{\scalerel*{
\begin{tikzpicture}[yscale=-1,transform shape]
\pic{orcidlogo};
\end{tikzpicture}
}{|}}}}

\usepackage{hyperref}

\title[Outlying E+A Galaxies in GAMA]{Identification of Anomalous E+A Galaxies in GAMA Using an Isolation Forest}

\author[K. Broadbelt et al.]{
Kieran Broadbelt$^{1}$ \orcidicon{0009-0001-6333-3270},
Kevin Pimbblet$^{1,2}$ \orcidicon{0000-0002-3963-3919},
and Daniel J. Farrow$^{1,2}$ \orcidicon{0000-0003-2575-0652}
\\
$^{1}$E.A. Milne Centre, Faculty of Science and Engineering, University of Hull, Cottingham Road, Hull HU6 7RX, UK\\
$^{2}$Centre of Excellence for Data Science, Artificial Intelligence and Modelling (DAIM), University of Hull, Cottingham Road, Hull HU6 7RX, UK
}

\date{Accepted XXX. Received YYY; in original form ZZZ}

\pubyear{\the\year{}}

\begin{document}
\label{firstpage}
\pagerange{\pageref{firstpage}--\pageref{lastpage}}
\maketitle

\begin{abstract}
We implement an outlier detection model, an Isolation Forest (iForest), to uncover anomalous objects in the Galaxy and Mass Assembly Fourth Data Release (GAMA DR4). The iForest algorithm is an unsupervised Machine Learning (ML) technique. The data we use is spectroscopic and photometric data from GAMA DR4, which compiles information for over 300,000 objects.  We select two samples of galaxies to isolate, high signal-to-noise galaxies, to analyse the iForest's robustness, and E+A galaxies, to study the extremes of their population. We result in six sub-samples of spectroscopic, photometric and combined data isolations, finding 101 anomalous objects, 50\% of which have not been identified as outliers in other works. We also find a number of fringing errors and false emission lines, displaying the iForest’s potential in detecting these errors. We find anomalous E+A galaxies - that although selected in a ‘normal’ manner using low $\rm{[OII]}$ and strong $\rm{H\delta}$ absorption - are still star-forming, with strong $\rm{H\alpha}$ emission. We propose two solutions to how these E+A galaxies are still star-forming but also question if these galaxies can be truly classified as E+A galaxies. We suggest that small-scale interactions of gas poor objects cause small star bursts, but the radiative pressure when high mass star form, expels the accreting material quicker than it can be accreted. We also suggest that the Jeans limit in our anomalous E+A galaxies is so low that it is simply not possible for O and B class stars to form, but it does not entirely prevent star formation.
\end{abstract}



\defcitealias{BaronPoz2016}{BP16}


\section{Introduction}
\label{sec:Intro}

As galaxy data expands rapidly with large surveys such as GAMA \citep{Driveretal2022}, Rubin LSST \citep{LSSTRubin}, DESI \citep{DESI}, Euclid \citep{Euclid}, we are left with an immense wealth of information to process. A current example of this wealth of data, GAMA DR4 \citep{Driveretal2022}, compiles over 300,000 spectra and results in 253,144 reliable galaxy redshifts. Furthermore, the Sloan Digital Sky Survey (SDSS) has also provided the community with multi-coloured images of more than three million objects \citep{Yorketal2000}. These immense datasets have led to a wide range of automated tools that can characterise and classify galaxies \citep{BaronPoz2016, Lochneretal2016, Clarketal2020, Reza2021, Changetal2021}. But, we still need more tools to probe for novel objects that can provide new information. ML algorithms are already in use to identify pre-defined objects like supernovae, irregular galaxies, active galactic nuclei (AGN), etc. \citep{Lochneretal2016, Reza2021, Changetal2021} and in use to classify previously unseen data into known classes \citep{Clarketal2020}. These algorithms, although effective at identifying known classes of object, do not extract the novel, anomalous objects we believe can provide new insights for astronomers.

Because of the size of astronomical datasets it is impossible for researchers to physically view and study all of the stars, galaxies, artifacts and data that is appearing. One such answer for this problem comes from outlier detection models \citep{IsolationForest, AmitMichele2015, BaronPoz2016, Margalefetal2020}. Outlier detection will reveal new information that can provide insight into current questions that astronomers have. \citet{BaronPoz2016} (hereafter \citetalias{BaronPoz2016}) uses an unsupervised Random Forest (RF) algorithm on the over two million spectra in SDSS and find 400 anomalous galaxies that are further analysed and studied. Some of these objects are AGN galaxies, post-starburst galaxies, extreme starformers and more. Their unsupervised ML algorithm identifies a large number of objects that are not found through other ML techniques. \citetalias{BaronPoz2016} also find errors in the SDSS pipeline, wrong classifications, `bad' spectra, and unusually broad [OIII] emission lines.

The main population of unusual galaxies that we aim to study are post-starburst galaxies. These galaxies are an important link between the star-forming spiral galaxies and their evolution to quiescent elliptical/S0 galaxies \citep{DressGunn1982, CouchSharp1987, Wilkinsonetal2017}. Post-starburst galaxies can also help in the understanding of how the environment influences the evolution of galaxies \citep{DressGunn1983, Zabludoffetal1996}. One type of post-starburst galaxy in particular, is thought to be a transitionary phase between star-forming spirals and quiescent elliptical galaxies, namely, E+A galaxies. E+A galaxies were identified by \citet{DressGunn1983} and have elliptical morphology whilst having high populations of A-class stars. These E+A galaxies display spectra with no $\rm{[OII]}$ emission but have deep Balmer ($\rm{H\delta}$) absorption lines. Strong $\rm{[OII]}$ is an indicator of ongoing star formation whilst the deep Balmer lines is a sign of young, recently formed, A-class stars \citep{Wilkinsonetal2017}. There is a variance however on how these E+A galaxies are defined, with some observational work selecting low $\rm{[OII]}$ emission and strong H$\delta$ absorption \citep{Poggiantietal2009, Verganietal2010, Wilkinsonetal2017}. Other authors however, use a lack of $\rm{H\alpha}$ emission in their selection \citep{Hoggetal2006, Goto2007a, Wilkinsonetal2017, Chen2019, Greene2021} to find `pure' E+As. \citet{Wilkinsonetal2017} compares these selection techniques, as does \citet{Greene2021}, and more detail about how our selection is performed will be discussed in \ref{sec:PSB}. The results of \citet{Wilkinsonetal2017} shows that the cuts of low $\rm{[OII]}$ and strong H$\delta$ find a population of green valley discs, a middle step between star-forming spirals and quiescent elliptical galaxies. The additional cut on low H$\alpha$ emission shows mostly early-type red galaxies and are classified as `pure' E+As.

In this work, an iForest \citep{IsolationForest} algorithm will be used on the spectroscopic and photometric data of our samples. The iForest is a novel approach at outlier detection that specifically targets outlying instances rather than profiling the normal instances. We will test the iForest on a sample of high S/N galaxies by applying a cut on the S/N measure. The S/N sample will be a secondary sample to the E+A sample and will consist of nearly 10,000 galaxies from GAMA that have S/N$\ge8$. 


\section{Data}
\label{sec:Data}

We take our data from GAMA DR4 \citep{Driveretal2022} and the necessary data management units (DMU) required for analysis \citep{Liskeetal2015, Gordonetal2017, Bellstedtetal2020}. As mentioned in section \ref{sec:Intro}, GAMA DR4 measured over 250,000 redshifts, and, in combination with earlier surveys, results in over 300,000 spectra across five sky regions \citep{Driveretal2022}. GAMA also catalogues a large number of DMUs that contain a wide variety of additional galaxy properties. The DMUs we use are: {\sc{speclinessfrv05}} \citep{Gordonetal2017} that compiles the line flux and equivalent width (EW) measures for the GAMA spectra; {\sc{speccatv27}} \citep{Liskeetal2015} that contains the spectra images, fibre data and other observation information; {\sc{apmatchedcatv06}} \citep{Liskeetal2015} that contains photometric data for the SDSS $urgiz$ and VIKING $ZYJHK$ surveys; and {\sc{gkvinputcatv02}} \citep{Bellstedtetal2020} that contains the photometric data in $FUV$, $NUV$, $ugri$, $ZYJHK$, $W1$ and $W2$ bands. For the spectroscopic analysis, we isolate only the EW measures, of which there are 51 features. For the photometric isolation, we isolate on the features of flux, magnitude and kcorrected colours, the calculations of which are discussed below, resulting in 326 features. The exact waves lengths, fitting procedures and other information about these values can be found in \citet{Gordonetal2017} for the spectroscopic measures and \citet{Liskeetal2015} and \citet{Bellstedtetal2020} for the photometric measures. For our analysis we have directly taken values from the aforementioned DMUs with no additional scaling or normalization.

We utilise colour in our photometric isolations but there is no direct colour DMU in GAMA DR4. We take the magnitude values from {\sc{apmatchedcatv06}} and use the Python module {\sc{kcorrect}}\footnote{\hyperlink{https://kcorrect.readthedocs.io/en/stable/index.html}{https://kcorrect.readthedocs.io/en/stable/index.html}} to scale the magnitudes to $z=0$. We then calculate the colours for the SDSS $ugriz$ bands and the VIKING $ZYJHK$ bands and input them back in to the feature list alongside the magnitude and flux values for future isolation.

With outlier detection in mind, we introduce a quality selection on the redshift quality of nQ $\ge$ 3 which reduces the data to approximately 230,000 spectra. The GAMA DMU {\sc{speclinessfr}} uses the value nQ to represent redshift quality. This quality selection is to ensure our outliers are not simply noisy or messy data and it confirms that the redshift is secure \citep{Driveretal2022}. \citet{Gordonetal2017} also introduce `dummy' values in to the spectroscopic measures. These dummy values occur where there were either: i) not enough pixels to perform a fit, ii) in the case of direct summation lines, they are in magnitudes and so logarithm of the flux returns an invalid value or iii) the flux value for molecular lines is always set at -99999.0, \citep{Gordonetal2017}. These dummy values would corrupt the iForest algorithm, so they are replaced with the mean of the non-dummy values in the column. This could cause false flags or incorrect outliers but we believe it to be better than simply removing all objects with dummy values.

The image data used for later morphological analysis comes from the SDSS survey, extracted via the {\sc{astroquery}} Python package \citep{Astroquery}. The SDSS survey contains approximately 3 million multi-coloured images \citep{Yorketal2000} for use in analysis. We use these images alongside a second Python package {\sc{statmorph}} \citep{RodriquezGomez} that has been designed and tested for calculating morphometric parameters.


\section{Methodology}
\label{method}
Here we will introduce in more detail the iForest being utilised in section \ref{sec:IForest} and two main samples we are applying it to, the E+A sample and the S/N sample.  Lastly, we will discuss the morphometric parameters that will be used to support the classification of our anomalous galaxies in section \ref{sec:MorphParam}.

\begin{figure*}
    \includegraphics[width=\textwidth]{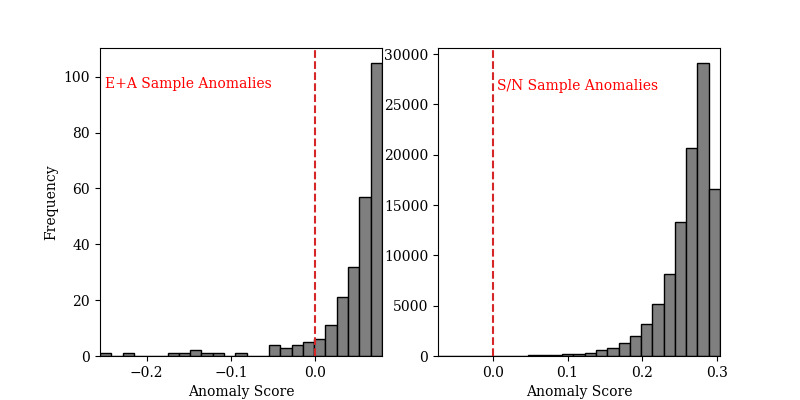}
    \caption{Anomaly scores of the two samples. Left: High S/N sample; consisting of $\sim10^5$ objects, the red dashed line delimits where anomalies are, below 0 indicates anomaly, above indicates nominal. Right: E+A sample; consisting of 287 objects, with the same delimiting line as the S/N plot.}
    \label{fig:AnomalyScores}
\end{figure*}

\subsection{Isolation Forest}
\label{sec:IForest} 

Anomalies are defined as data patterns that have different characteristics from normal instances \citep{IsolationForest}. The detection of these anomalies can be of significant importance when discovering novel data that can be probed for new information. In section \ref{sec:Intro}, most of the existing model-based approaches to outlier detection will construct a profile of the standard instances within the data and will then identify instances that do not fit that profile. Notable examples in astronomy outlier detection include the use of Deep Generative Networks \citep{Margalefetal2020}, statistical and machine learning techniques \citep{AmitMichele2015}, and RF algorithms (\citetalias{BaronPoz2016}). Many of these existing methods are constrained to low dimensional data because of their high computational complexity \citep{IsolationForest, TahiHadi2019}. This issue is something that recent literature has been attempting to overcome \citep{Liuetal2018, Kamalov2019}.

\citet{IsolationForest} propose a different model-based method that explicitly isolates outlying data, rather than profiling normal instances. This is achieved by utilising two quantitative properties of anomalous data: i) anomalies will be fewer than normal instances and ii) they have attribute-values that are especially different from those of normal instances \citep{IsolationForest}. What this means is that anomalies by definition are going to be few and different, meaning, they are more susceptible to isolation. Using this idea, a tree structure is constructed that isolates every instance in the dataset, called an Isolation Tree (iTree). Because of the fact anomalies are more susceptible to isolation, they are more likely to be found at the base of the iTree structure, whereas normal instances would be found in the deeper `branches'. Due to the outlying data lying close to the root of the tree, the largest part of the tree, that isolates the normal instances, does not need to be generated. This allows the model to build partial models and to exploit sub-sampling \citep{IsolationForest}.

Sub-sampling is a necessary step in the process of the iForest method. Large sample sizes can reduce the iTree's ability to confidently identify anomalies and thus the iForest as a whole becomes less effective. This is due to swamping and masking. Swamping is when the normal instances are too similar to the outlying instances, and masking is where the existence of too many outliers hides their own presence \citep{IsolationForest}. Sub-sampling overcomes these issues by creating smaller, partial selections of the data. This leads to smaller iTrees that can better isolate data, as well as each iTree being able to specialise as each sub-sample would contain different sets of anomalies or even no anomalies. Furthermore, this allows the algorithm to be robust to overfitting and avoids the need for expensive distance calculations with larger iTrees \citep{Ishida2021}.

The iForest in this work is trained on two key samples. The samples are E+A galaxies (see section \ref{sec:PSB}) and high S/N galaxies (see section \ref{sec:Weird}). The data given to the iForest includes both spectroscopic data (equivalent width (EW) measures) and photometric data (colour, mag and fluxes). Three isolations are run on the galaxy selections. The first iForest is given purely the spectroscopic data, the second iForest is given the photometric data and the third is provided with both spectroscopic and photometric data. The iForest will use a sub-sample sizes of 256 instances to manage this data which is the standard value in \citet{IsolationForest}. Figure \ref{fig:AnomalyScores} displays the resultant anomaly scores of the spectroscopic isolations of both samples. As stated, a score below zero (the red dashed line) indicates an anomalous object and we selected to isolate 25 anomalies from each sample.


\subsection{E+A Galaxies}
\label{sec:PSB}
E+A galaxies are a strong tool for investigating the processes of galaxy evolution. They are an important link between star-forming spirals and quiescent ellipticals \citep{DressGunn1982, CouchSharp1987, Wilkinsonetal2017}. We discussed the many selection methods in section \ref{sec:Intro} and there are merits and pitfalls of each method. More recent work have stated the necessity of limiting $\rm{H\alpha}$ \citep{Chen2019, Greene2021}. $\rm{H\alpha}$ is an indicator of star formation and although this selection can remove dusty star formers, it can remove E+A galaxies that have $\rm{H\alpha}$ because of AGN activity \citep{Wilkinsonetal2017}. This AGN possibility is something we come across in our later analysis of our results (section \ref{sec:psbresults}). We also wish to further study the selection method and the homogeneity of such selection methods.

\citet{Wilkinsonetal2017} studies three selections in depth in their work. They have an additional look at solely strong H$\delta$ Balmer absorption to further analyse the selection process. The master sample of H$\delta$ strong galaxies uses an EW$_{\rm{H\delta}} > 3$\r{A}. From this sample, a further two selections are made, both have low $\rm{[OII]}$ emission using EW$_{\rm{[OII]}} > -2.5$\r{A} called the `E+A' sample. The third selection has a final cut of low $\rm{H\alpha}$, EW$_{\rm{H\alpha}} > -3$\r{A} called the `pure E+A' sample. \citet{Wilkinsonetal2017} concludes that the $\rm{H\delta}$ strong sample commonly have late-type morphology and are often star formers. The E+A sample have a mix of elliptical and spiral morphologies. Lastly, the pure E+As are most commonly red elliptical galaxies and have a higher fraction in denser environments compared to the H$\delta$ strong and E+A galaxies. \citet{Wilkinsonetal2017} states that these results suggest an evolutionary sequence from blue-disc galaxies to quiescent red elliptical galaxies.

We want to study the evolution of the E+As into the quiescent early types and as such, choose to use the E+A selection \citep{Wilkinsonetal2017}. As such we use H$\delta$ strong as EW$_{\rm{H\delta}} < -3$\r{A}, and low [OII] absorption as EW$_{\rm{[OII]}} < 2.5$\r{A}. We change the signs to fit with the GAMA conventions for denoting emission and absorption lines. This selection keeps the $H\alpha$ strong galaxies in the sample and will allow for future analysis of AGN activity in the galaxy. AGN-driven feedback is thought to be a possible mechanism for quenching galaxies e.g. \citep{Hopkinsetal2008, Wilkinsonetal2017}.

We apply additional S/N cuts to the EW measures. This is to ensure that we are not simply finding noisy galaxies that might muddy the results. The S/N cut is performed as follows,
\begin{equation}
    \rm{S/N} = \frac{\rm{|EW|}}{\rm{{EW}_{err}}} \ge 3. 
\end{equation}
Through our selection method for E+A galaxies we create a sample of 287 galaxies. This sample constitutes $\sim$1\% of the GAMA DR4 total dataset, a typical amount when compared to the local Universe \citep{Greene2021}.


\subsection{High S/N Galaxies}
\label{sec:Weird}
The signal to noise cut is designed to remove the noisiest data from the sample and leave only high S/N galaxies. We implement a simple S/N > 8 cut to the full 230,000 spectra in the sample, resulting in a sample of $\sim10^5$. Similar to the E+A galaxy isolation, we perform the iForest with three inputs of data, spectroscopic, photometric and a combination of both. This sample is secondary to the E+A sample and is here as test of the iForest's ability to isolate anomalies in a larger sample.


\subsection{Morphometric Parameters}
\label{sec:MorphParam}
We will be using 8 morphometric parameters. Concentration, asymmetry and smoothness \citep[$CAS$][]{Conselice2003, Bershadyetal2000}, Gini and $M_{20}$ \citep{Abraham2003, Lotzetal2004} are used extensively in optical images of galaxies and are a versatile tool in quantising the shape, size and other morphological qualities of galaxies. We will also include a less tested but interesting set of morphometry parameters: multimode, intensity and deviation \citep[$MID$][]{Freeman2013, Peth2015}. We utilise {\sc{statmorph}} \citep{RodriquezGomez}, a commonly used Python tool, to compute these morphometrics using the standard definitions from the above mentioned papers. 


\subsubsection{Concentration-Asymmetry-Smoothness (CAS)}
\label{sec:CAS}
CAS is a commonly used space that was developed by \citet{Bershadyetal2000} and \citet{Conselice2003}.
Concentration ($C$) measures the ratio of light within an inner aperture to the light within an outer aperture. The definition in \citet{Bershadyetal2000} states,
    \begin{equation}
        \label{eq:concentration}
        C = 5\log_{10}\left(\frac{r_{80}}{r_{20}}\right),
    \end{equation}
where $r_{20}$ and $r_{80}$ are the radii of the circular apertures that contain 20\% and 80\% of the total flux respectively. We take the total flux to be the flux contained within 1.5 times the Petrosian radius ($r_p$), the standard definition from \citet{Conselice2003}. The centre for these apertures is determined via the minimisation of asymmetry which we will discuss below.
High $C$ values indicate a bright central bulge region which is a main feature of spiral galaxies. The circular apertures about the centre however, make this measure inconsistent if the bright structures exist outside this central region, e.g multiple nuclei. Typical values for discs are $C > 3.5$ and for large ellipitcals, $C = 1 - 3$. Irregular galaxies span the entire spectra \citep{Conselice2003}


Asymmetry ($A$) is defined as the number computed when a galaxy is rotated $180^{\circ}$ about its centre and then subtracted from the original galaxy image \citep{Conselice2003}:
    \begin{equation}
        \label{eq:asymmetry}
        A = \sum _{ij} \frac{|I(i,j) - I_{180}(i,j)|}{|I(i,j)|} - B_{\rm{asymm}},
    \end{equation}
where, $I$ is the galaxy's image and $I_{180}$ is the rotated image, $i$ and $j$ describe the pixel positions on a 2D image. $B_{\rm{asymm}}$ is an estimate of the contribution the sky background has on the asymmetry measure. This means $A$ can be negative if the background asymmetry is large.  Due to the noise correction, it is unreliable to compute $A$ for low S/N images, this should not be an issue in our data as we have set lower limits to S/N before calculations. Like $C$, $A$ is calculated within 1.5$r_p$ about the galactic centre, this centre is determined by minimising $A$. This minimisation is calculated by moving the centre of rotation about a grid at the centre of the image. A high asymmetry value indicates an asymmetric galaxy, typical of irregular and merging galaxies. Smooth, elliptical light profiles will result in a lower $A$ value due to their rotational symmetry.


Smoothness ($S$) is the final parameter developed by \citet{Conselice2003}, utilised to quantify the degree of small-scale structure. The image of the galaxy is smoothed and then subtracted from the original image:
    \begin{equation}
        \label{eq:smoothness}
        S = \sum _{ij} \frac{|I(i,j) - I_{\rm{smooth}}(i,j)|}{|I(i,j)|} - B_{\rm{smooth}}.
    \end{equation}
Here, $I_{\rm{smooth}}$ is the smoothed galaxy image and $B_{\rm{smooth}}$ is an estimate of the contribution the sky background has on the smoothness measure. In this work we use a gaussian smoothing with a standard deviation of 0.25$r_p$. $S$ is typically an indicator of recent star-formation as the small-scale structures are often where star-formation is occurring in late-type spirals. Smooth ellipticals that are no longer star forming will have a low $S$ value \citep{Conselice2003}. Like $A$, $S$ is not as effective when applied to poorly resolved galaxies such as those at high redshifts.


\subsubsection{Gini and $M_{20}$}
\label{sec:gini}
\citet{Abraham2003} and \citet{Lotzetal2004} introduce the $GM_{20}$ space as an alternate measure to $CAS$. Their aim was to create parameters that are free from the centre selection processes and are more susceptive to merger and interaction signatures.

The Gini coefficient ($G$) is a statistical measure based on the Lorenz curve, the rank-ordered cumulative distribution function of a population's wealth. A Gini value of 0 indicates perfect equality (all pixels have an equal fraction of the flux) and a Gini value of 1 indicates perfect inequality (a single pixel contains all the flux). This measure is repurposed as a distribution measure of the galaxy's pixel values \citep{Abraham2003, Lotzetal2004}. For a discrete population, $G$ can be defined as the mean of the absolute difference between all pixel values. \citet{Lotzetal2004} sort the pixel values into increasing order to allow for more efficient computation and use the definition,
    \begin{equation}
        \label{eq:giniindex}
        G = \frac{1}{|X|n(n-1)}\sum_i^n(2i-n-1)|X_i|,
    \end{equation}
where $X$ is the mean over all pixel flux values $X_i$, $n$ is the total number of pixels in the galaxy, $i$ is the pixel index.
For the majority of local galaxies, $G$ correlates with $C$ and will increase with the ratio of light in the central component of the galaxy. \citet{Abraham2003} finds that $G$ strongly correlates with surface brightness and $C$ when studying 930 galaxies in the SDSS Early Data Release. $C$ and $G$ deviate however because $G$ is found independently from the spatial distribution of flux. Strong $G$ values can be found when bright pixels exist outside of the central component, where as strong $C$ values typically denote only a bright central component \citep{Conselice2003, Lotzetal2004}. When computing $G$, care must be given to the background sky. $G$ must be calculated from only the pixels belonging to the galaxy in order to be a true and accurate measurement of the galaxy's $G$ coefficient \citep{Lotzetal2004}. 


The second-order moment of the brightest 20\% of the flux ($M_{20}$) is another measure of flux distribution. We first define the second order moment of a pixel $i$ as in \citet{Lotzetal2004},
    \begin{equation}
        \label{eq:mi}
        M_i = I_i[(x-x_c)^2 + (y-y_c)^2].
    \end{equation}
Where $x,y$ is the position of a pixel with intensity value $I_i$ in the image and $x_c,y_c$ is the central pixel position of the galaxy in the image. The total second order moment is then given by:
    \begin{equation}
        \label{eq:mtot}
        M_{\rm{tot}} = \sum_i^nM_i.
    \end{equation}
\citet{Lotzetal2004} uses the relative contribution to the second order moment of the pixels that contain 20\% of the total flux  after sorting the list of pixels by descending intensity, giving us:
    
    \begin{equation}
        \label{eq:m20}
        M_{20} = \log_{10}\left(\frac{\sum_iM_i}{M_{\rm{tot}}}\right) \text{ while } \sum_iI_i<0.2I_{\rm{tot}}.
    \end{equation}
Here $I_{\rm{tot}}$ is the total flux of the galaxy pixels. $M_{20}$ is sensitive to bright regions in the outer regions of discs and a larger value for $M_{20}$ indicates star-forming outer regions or strongly interacting discs. Furthermore, $M_{20}$ is more sensitive to multiple nuclei than $C$ as it does not use apertures in its implementation and its centre point is a free parameter.


\subsubsection{Multimode-Intensity-Deiviation (MID)}

The MID statistics \citep{Freeman2013, Peth2015} were first introduced as an alternative to the $CAS$ and $GM_{20}$ parameters.  Their aim is to be more sensitive to recent mergers than the prior morphometrics and to overcome the difficulties in identifying post-merger morphologies. They also aimed to overcome the degradation of $CAS$ and $GM_{20}$ that comes with increasing redshift. These statistics have not been tested as extensively as the $CAS$ and $GM_{20}$ measures, especially using hydronamical simulations \citep[see discussion in][]{RodriquezGomez, Holwerda2025}.

The multimode $M$ statistic  measures the ratio between the areas of the two most `prominent' clumps within a galaxy. This has the implicit assumption that a well resolved galaxy will have at least two well-defined clumps \citep{Holwerda2025}. The bright regions are found using a threshold method where $q_l$ represents the normalized flux value and $l$ percent of pixel fluxes are less than $q_l$. This results in a binary image $g_{i,j}$ where 1 represents fluxes larger than $q_l$ and 0 represents fluxes less than $q_l$:
    \begin{equation}
    g_{i,j} = \begin{cases}
    1 & f_{i,j} \ge q_l\\
    0 & \text{otherwise}
    \end{cases}
    \end{equation}
The number of pixels in contiguous groups of pixels with value 1 are then sorted in descending order by area. The two largest groups $A_{l,(2)}$ and $A_{l,(1)}$ define an area ratio $R_l$ \citep{Peth2015}:
    \begin{equation}
        R_l = \frac{A_{l,(2)}}{A_{l,(1)}}.
    \end{equation}
The $M$ statistic is then the maximum value of $R_l$. Values of $M$ that approach 1 represent multiple nuclei, while values near 0 are single nuclei systems \citep{Peth2015}. The original work of \citet{Freeman2013} applies an additional factor of $A_{l,(2)}$ to limit the affects of hot pixels, but {\sc{statmorph}} utilises the above method from \citet{Peth2015} so the measure is size independent.

The intensity statistic ($I$) measures the ratio between the two brightest subregions of the galaxy. To calculate $I$, the galaxy image must first be smoothed slightly using a Gaussian kernal with $\sigma = 1$ pixel. The image is then partioned into pixel groups according to the watershed algorithm. This algorithm states that each distinct subregion consists of all the pixels such that their maximum gradient paths lead to the same local maximum. The surrounding eight pixels of every pixel are inspected and the path of maximal intensity increase is followed until a local maximum is reached. These maximums are the local brightness maximums of the pixels flux \citep{RodriquezGomez}. Once the pixel groups are defined, their summed intensities are sorted into descending order: $I_1, I_2, etc.$ Intensity is then defined as \citep{Freeman2013}:

\begin{equation}
    I = \frac{I_2}{I_1}.
\end{equation}

Lastly, Deviation ($D$) measures the distance between the intensity centroid ($x_{\rm{c}},y_{\rm{c}}$), calculated for the pixels identified by the $MID$ segmentation map, and the brightest peak found during computation of the $I$ statistic \citep[$x_{I_1},y_{I_1}$][]{Freeman2013}. The intensity centroid is defined as,
    \begin{equation}
        (x_{\rm{c}},y_{\rm{c}}) = \left( \frac{1}{n_{\rm{seg}}}\sum_i\sum_jif_{i,j}, \frac{1}{n_{\rm{seg}}}\sum_i\sum_jjf_{i,j}\right).
    \end{equation}
with the summation being overall $n_{\rm{seg}}$ pixels within the segmentation map. The distance between the two centroids will be affected by the absolute size of the galaxy and as such a normalisation is applied using $\sqrt{n_{\rm{seg}}/\pi}$. The $D$ statistic is the defined as:
    \begin{equation}
        D = \sqrt{\frac{\pi}{n_{\rm{seg}}}}\sqrt{(x_{\rm{c}}-x_{I_1})^2 +(y_{\rm{c}}-y_{I_1})^2}.
    \end{equation}


\begin{figure*}
    \centering
	\includegraphics[trim={120 120 120 120}, clip, width=\textwidth]{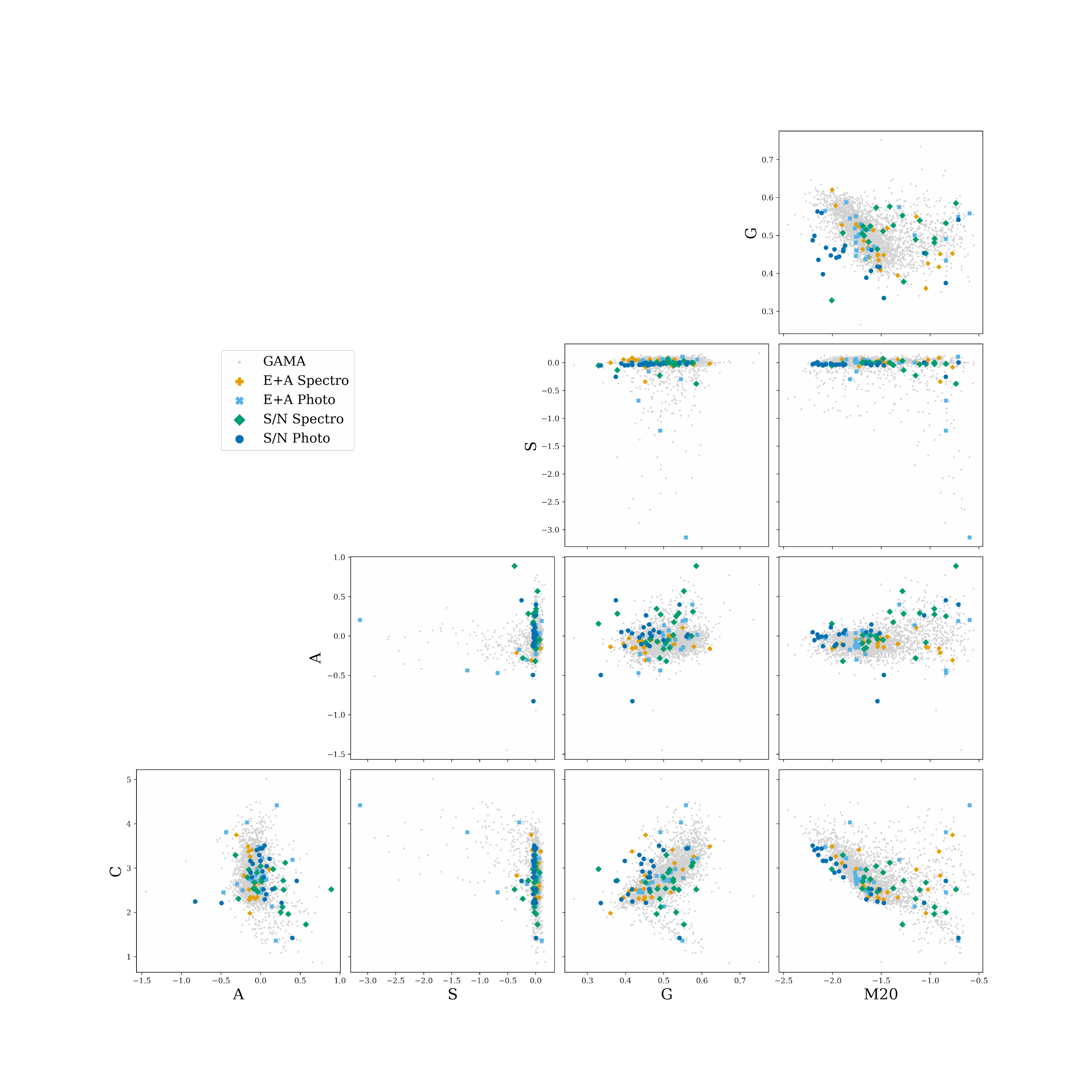}
    \caption{A corner plot of the most common 5 morphometric measures, concentration, asymmetry, smoothness, Gini \& $M_{20}$. Orange pluses are E+A spectroscopic anomalies, light blue crosses are the E+A photometric anomalies, green diamonds are S/N spectroscopic anomalies and navy blue circles are S/N photometric anomalies. We can see from this corner plot that a significant portion of the anomalous objects, have `normal' morphologies, in that, they lie within the bulk of the other objects in GAMA. There are some extreme outliers morphologically such as in asymmetry and smoothness.}
    \label{fig:CASGM}
\end{figure*}


\subsubsection{Automated Morphology Measuring}
\label{sec:statmorph}

We supply {\sc{statmorph}} with cutout images extracted from SDSS using {\sc{astroquery}} \citep{Astroquery}. The initial cutout size of each image is a 50px by 50px square, which provides 20 by 20 arcsecond image. We then apply a threshold mask to find the objects in the cutout and reduce the cutout size of the galaxy to 1.5$r_p$ of the galaxies flux. By supplying these images to {\sc{statmorph}}, it can extract the full morphometric space and give us our 8 parameters and more \citep[see in][]{RodriquezGomez}. Figure \ref{fig:CASGM} shows a corner plot of the most commonly used morphometric parameters, C,A,S,G \& $M_{20}$.


\begin{figure*}
\begin{centering}
    \includegraphics[width=13cm]{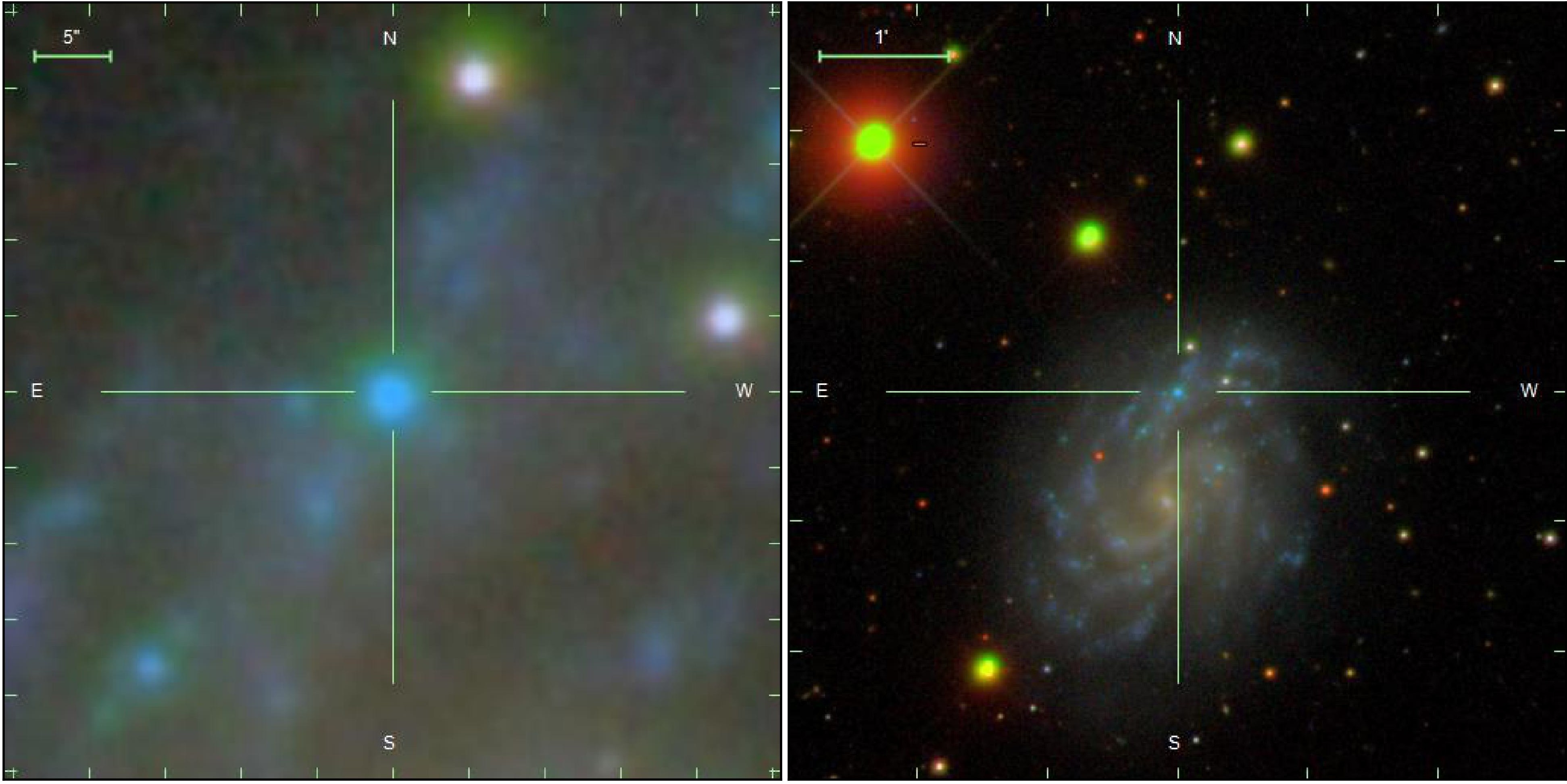}
    \caption{Example shredded galaxy. Left: The falsely identified galaxy that is actually a bright sub-structure in a larger galaxy, right.}
    \label{fig:Stripped}
\end{centering}
\end{figure*}

\begin{figure*}
\begin{centering}
    \includegraphics[trim={1.5cm 4cm 1.5cm 3cm}, clip, width=13cm]{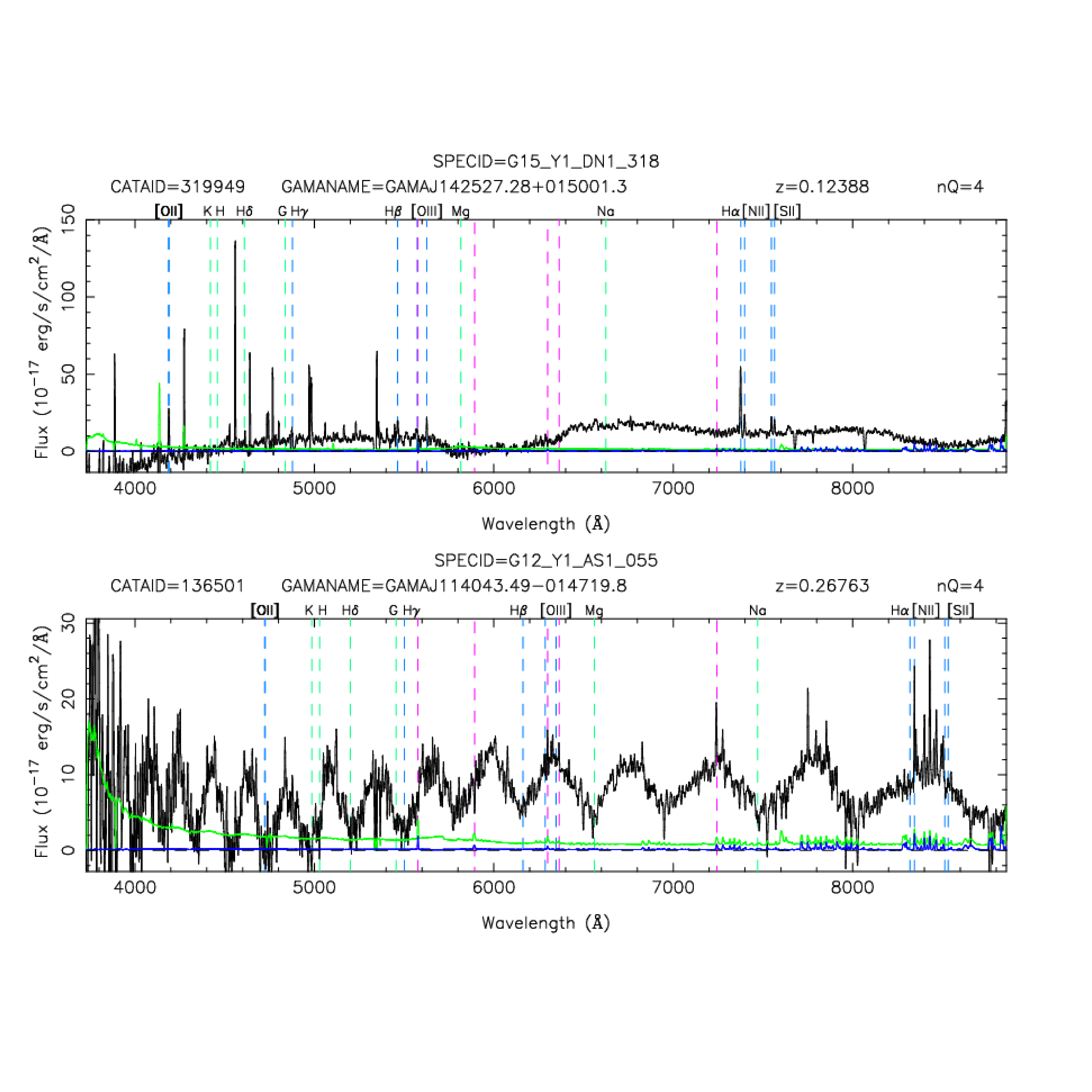}
    \caption{Two types of `bad spectra' found by the iForest. Top: false emission line spectra; displaying peaks in areas that are not valid emission lines. This is caused due to poor flat fielding that was an issue with early AAOmega that has since been fixed (Croom, priv. comm.). Bottom: data reduction error; this presents as a sinusoidal pattern in the spectra line.}
    \label{fig:falseemission}
\end{centering}
\end{figure*}

\begin{figure*}
\begin{centering}
    \includegraphics[trim={1.5cm 4cm 1.5cm 3cm}, clip, width=13cm]{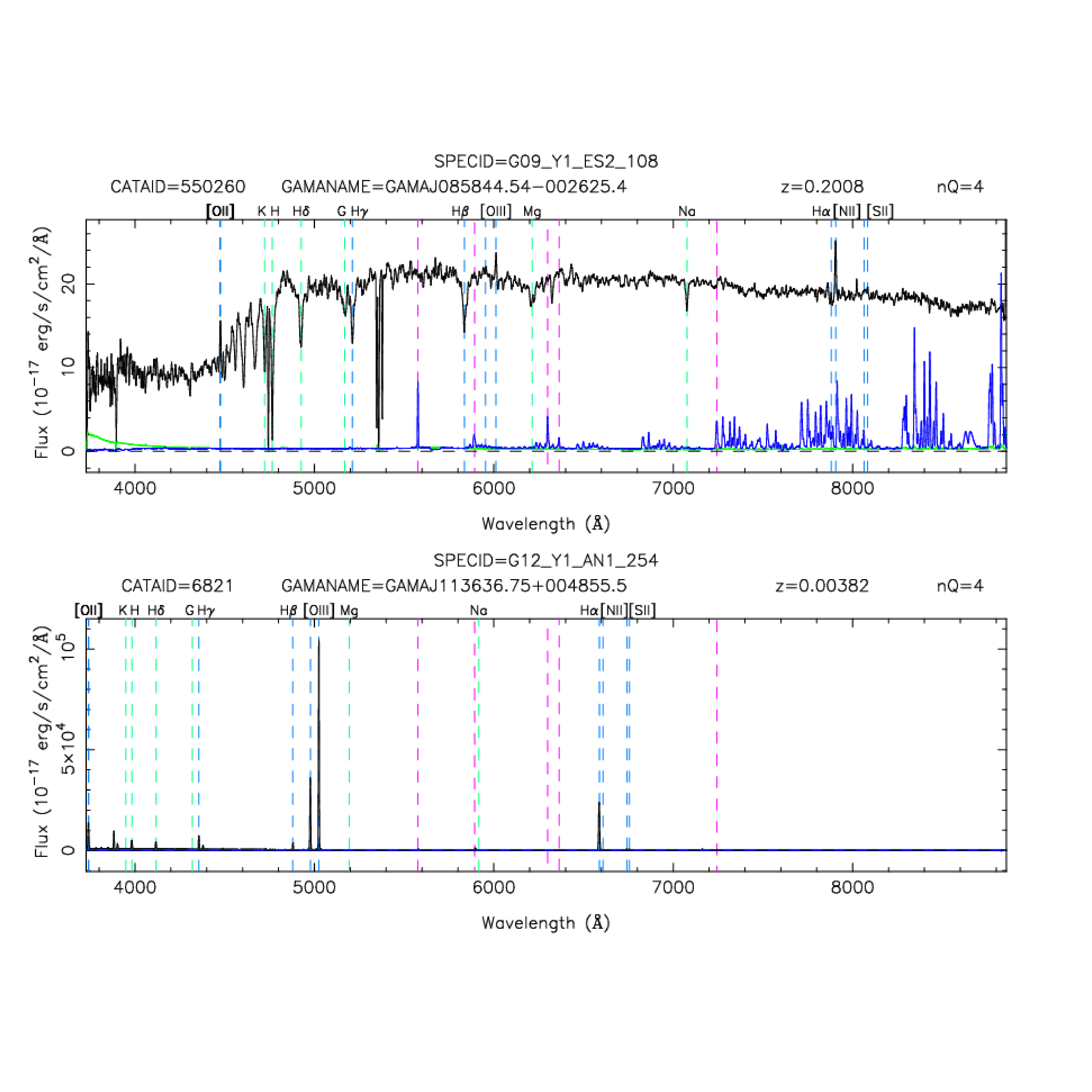}
    \caption{Two types of anomalous spectra found by the iForest. Top: strong ${\rm{H\alpha}}$ E+A galaxy extracted from GAMA DR4. This galaxy has a $\rm{EW}_{H\alpha} \approx 21$\AA, $\rm{EW}_{[OII]} \approx 2$\AA and $\rm{EW}_{H\delta} \approx -3$\AA. Bottom: EELG extracted from GAMA DR4. This galaxy has a $\rm{EW}_{[OIII]} \approx546$\AA.}
    \label{fig:phenomspectra}
\end{centering}
\end{figure*}


\section{Results}
We calibrate the iForest to find the top 25 most anomalous galaxies from each sample and dataset. We chose 25 because it have a feasible sample size to analyse, while still maintaining a statistically useful sample count. These isolations results in a total of 150 anomalous objects identified from the GAMA DR4 data. In total, we find that there are 103 unique objects due to duplicate anomalies identified across the spectroscopic, photometric and combined isolations. We find that two of those 103 objects are falsely identified as galaxies by the GAMA pipeline and are in fact shredded larger galaxies, this can be seen in figure \ref{fig:Stripped}. The remaining 101 galaxies are confidently identified as galaxies and are further inspected. Of the 101, we find that 13 have `bad spectra', consisting of pipeline issues, fringing errors, data reduction errors and false emission lines. This shows the iForest's ability to robustly extract data errors in a given sample without training. We display examples of two types of bad spectra in figure \ref{fig:falseemission}, namely the fringing error and the false emission lines.

We find 54 unique objects from the S/N spectroscopic, photometric and combined samples, 2 of which are the falsely identified shredded galaxies, resulting in 52 confirmed galaxies. 20 objects are explicitly isolated from the spectroscopic data, 22 are isolated from the photometric sample and/or the combined sample and 10 are found exclusively in the combined spectro/photo sample (see in table \ref{tab:SNtable}). These galaxies encompass a wide range of astronomical phenomena such as extreme emission line galaxies (EELG), a subset of EELGs known as `Green Beans', AGN hosts and red spiral galaxies. We find that 13 galaxies are EELGs, selected as having either $\rm{EW_{[OIII]}}$ or $\rm{EW_{H\alpha}}$ > 300\text{\AA} \citep{LumbrerasCalle2021} and an example spectra of one of these EELGs is displayed in figure \ref{fig:phenomspectra}. 9 of the identified EELGs are not cited by any works targeting EELGs. One of the EELGs identified is the aforementioned `Green Bean' galaxy \citep[see in][]{PresSand2019}. We use BPT and WHAN diagrams seen in figure \ref{fig:bptwhan} and identify 4 possible AGN hosts, one of which being the Green Bean galaxy. The largest sample of anomalous galaxies are the red spiral galaxies, consisting of 18 out of the 52 anomalies found from the S/N samples. This value equates to approximately 35\% of the anomalies in this sample. This leaves 18 galaxies that are not immediately considered scientifically important galaxies and from study of their spectra we see that many have spectral lines that lie below the continuum.

We choose to isolate $\sim$10\% of the E+A sample, resulting in 25 objects from each sample, 75 objects total. We find that 25 anomalous objects are isolated from the spectroscopic data and 24 objects are isolated from the photometric data (see in table \ref{tab:EAtable}). We note that no unique galaxies are found exclusively in the combined isolation. From the 49 objects isolated from the E+A sample, we find that there are 29 objects with abnormally large $\rm{H\alpha}$ emission lines (EW$_{\rm{H\alpha}} > 10$\AA). 20 of these galaxies come from the spectroscopic isolation, and 9 come from the photometric isolation. We expected that the anomalous E+A galaxies isolated from the spectroscopic data would have strong $\rm{H\alpha}$ but we also detect strong emission in some of the E+A galaxies found from the photometric isolation. The remaining population of objects isolated do not appear to be scientifically interesting, giving credence to the notion that just because objects are statistically rare, does not mean they are automatically interesting. Of the 16 remaining photometric anomalies, they all appear visually typical for E+A galaxies, being red in colour and elliptical in shape.


\section{Discussion}

\begin{figure}
\begin{centering}
    \includegraphics[trim={0cm 0.5cm 1.5cm 1.5cm}, clip,width=\columnwidth]{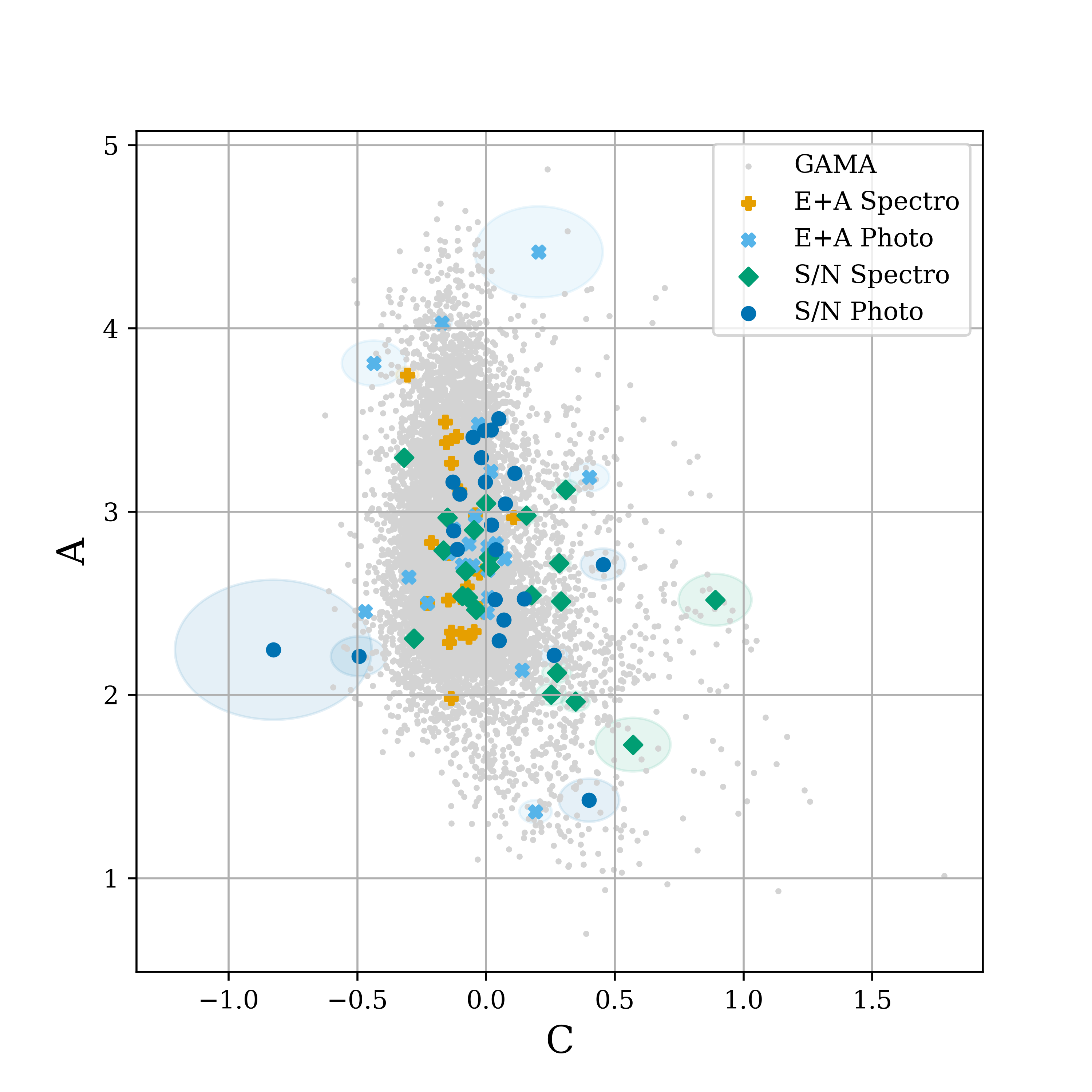}
    \caption{Concentration-Asymmetry plot used to find the population density of each object in the $C-A$ space. We find the distance to the nearest 10th neighbour to identify the anomalous objects with the most extreme morphology. The 9 galaxies with the lowest densities (most different) are displayed in figure \ref{fig:CASweird}.}
    \label{fig:casnearneighbour}
\end{centering}
\end{figure}

Using the 8 morphometric parameters discussed in section \ref{sec:MorphParam}, we can quantise the morphologies of our anomalies. Figure \ref{fig:CASGM} displays a corner plot of the parameters, concentration, asymmetry, smoothness, gini and M20. The plot indicates that the anomalous galaxies are typically normal in morphology. There are however some extremes, particularly within the smoothness and M20 regimes. We find the most anomalous galaxies by morphology, to do this we take the C-A plot and calculate the population density of the points near by, shown in Fig. \ref{fig:casnearneighbour}. From the population densities we show the 16 galaxies that have the lowest population density, and therefore are the most morphologically anomalous, in figure \ref{fig:CASweird}.

\begin{figure}
    \includegraphics[width=\columnwidth]{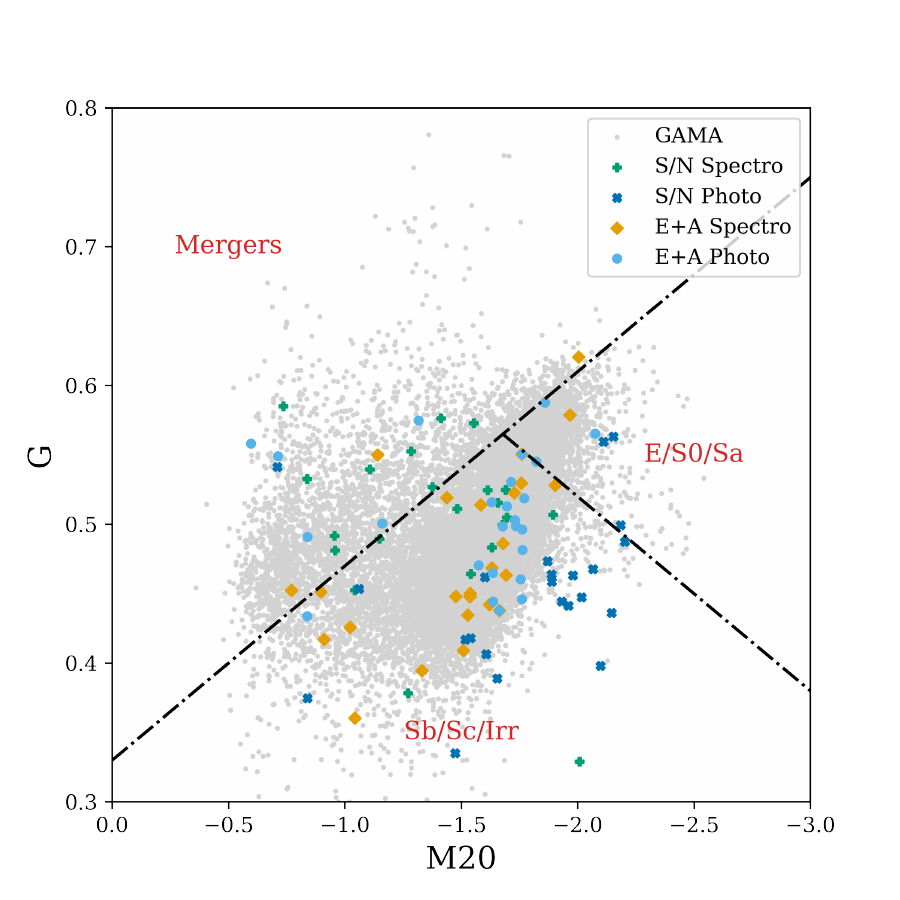}
    \caption{$G$-$M_{20}$ plot with delimiting lines from \citep{Lotzetal2008}. The key for the points are the same as in figure \ref{fig:CASGM}. We can see that the many of the E+A galaxies lie within the Sb/Sc/Irr region, where they should lie in the E/S0/Sa region. This suggests that something has disturbed the morphology of the E+A galaxies and could hint at a reason for the E+A galaxy's starburst.}
    \label{fig:GM20}
\end{figure}

The quantised parameters can be utilised to check if E+A galaxies follow the typical S0/elliptical shape. Using a Gini-$M_{20}$ plot and region lines discussed in \citet{Lotzetal2008} we roughly group our isolated objects into three regions, merger candidates, Sb/Sc/Irr candidates and E/S0/Sa candidates. From Fig. \ref{fig:GM20} we can see that the approximately 75\% of our isolated galaxies lie within the Sb/Sc/Irr region, this agrees with visual inspection as many of our isolated galaxies have irregular morphology. We can see that only 3 isolated E+A galaxies lie within the E/S0/Sa region. Since E/S0/Sa is the typical morphology of E+A galaxies \citep{DressGunn1983}, this again supports that the iForest algorithm has worked to find the weirdest E+A galaxies. We also find around about 20\% of the E+A galaxies are merger candidates, which could be the cause of their recent starbursts.

\begin{figure}
\begin{centering}
	\includegraphics[width=\columnwidth]{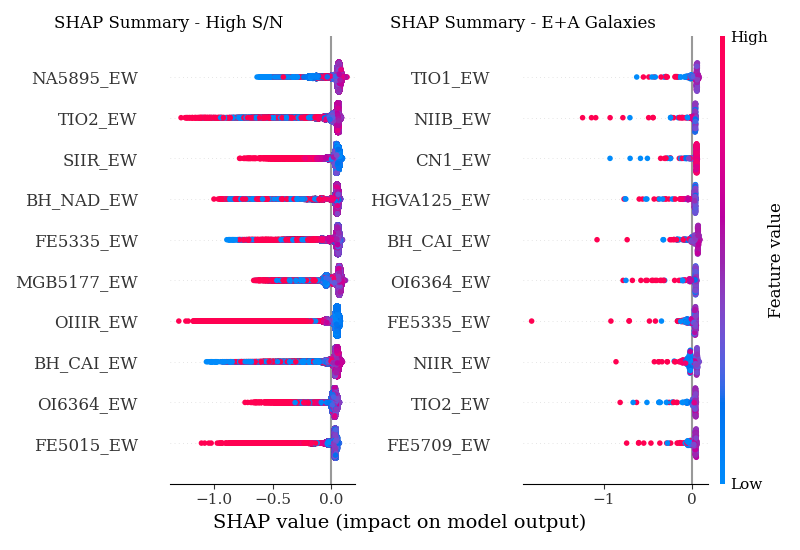}
    \caption{A summary plot created using the {\sc{shap}} Python tool. The plot shows the 10 most impactful features from the spectroscopic data set in descending order. Left: High S/N sample, we can see that some of the most impactful features are sodium and titanium, typical of small class stars like K and M. We also see [OIII] we is typical of extreme star-formers or AGN hosts. Right: E+A sample, we can again see titanium and iron, we also see nitrogen, hinting at possible AGN hosts.}
    \label{fig:shap_spectro}
\end{centering}
\end{figure}

We also utilise a Python tool {\sc{shap}} to analyse the impact value of each feature in the datasets. SHapley Additive exPlanations (SHAP) is a game theoretic approach to explain the output of any machine learning model \citep{shap}. It connects optimal credit allocation with local explanations using the classic Shapley values from game theory to quantise the `impact' each feature has on the model. From figure \ref{fig:shap_spectro} we can see that TiO2 and TiO1 has a strong impact on defining our anomalies, this line of TiO1/TiO2 suggests large populations of low mass stars. We can also see the forbidden line of $\rm{[OIII]}$ having a strong impact in the S/N sample. We are surprised that $\rm{H\alpha}$ is not in the top 10 most impactful features when studying the E+A spectroscopic sample, but, due the amount of features included in the isolation, each individual feature has a rather small impact compared to the overall dataset. Similarly, when considering the impact of the photometric features in figure \ref{fig:shap_photo}, each individual feature has a small impact due to the volume of data. We do note however that the u band photometry plays the largest role when designating anomalies for both the high S/N sample and the E+A sample.

\subsection{High S/N Anomalies}

\begin{figure*}
\begin{centering}
    \includegraphics[width=\textwidth]{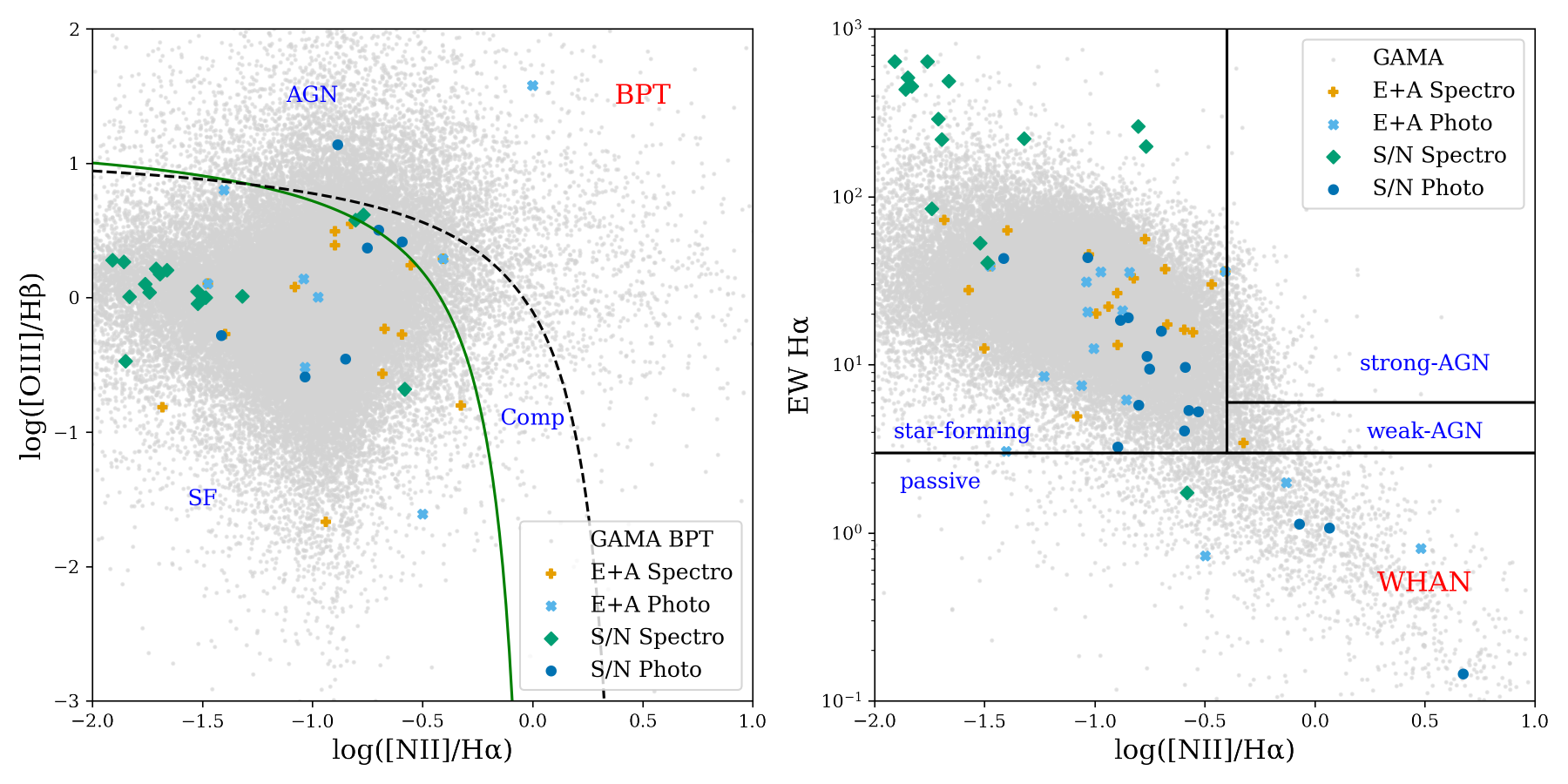}
    \caption{Left: BPT diagram; The black dashed line is the extreme starburst classifier line and the green solid line is the pure star formation line from \citet{Kewley2006}. Right: WHAN diagram; the lines delimit the spectral classes defined in \citet{WHAN}. The key is the same as in figure \ref{fig:CASGM}. We see 3 objects are possible AGN hosts from both BPT and WHAN diagrams. We also see that $\sim$85\% of the E+A galaxies lie within the star-forming region in the WHAN plot when they should lie in the passive region.}
    \label{fig:bptwhan}
\end{centering}
\end{figure*}

13 EELGs are isolated from the S/N sample, with either $\rm{EW_{[OIII]}}$ or $\rm{EW_{H\alpha}}$ > 300\text{\AA}. EELGs are a rare population of galaxies that have been studied extensively over the past decade \citep{Maseaetal2014, Smitetal2015, LumbrerasCalle2021, Llerenaetal2024}. These galaxies are especially rare in the local Universe, but their number density increases with increasing redshift \citep{Smitetal2015}. Typically, local EELGs have low stellar masses, $\rm{M_*<10^9M_\odot}$. The EELGs we have found have stellar masses ranging from  $\rm{10^8 \rightarrow 10^{11}M_\odot }$ and have redshift values $0.01<z<0.3$. The most common lines that we find EELGs in are $\rm{[OIII]}\lambda4959$\text{\AA}, $\lambda5007$\text{\AA} (here after as $\rm{[OIII]}$) and $\rm{H\alpha}$ in the rest-frame optical \citep{Iglesias2022, Llerenaetal2024}. We note that all 13 of the EELGs come from the spectroscopic isolation and find that 9 of these EELGs have not been identified by other works. One EELG that has been identified in other works is the `Green Bean' galaxy \citep{PresSand2019}. Green Beans are a similar galaxy to `Green Pea' galaxies, first noted in \citep{Cardamoneetal2009} and observed by volunteers in the Galaxy Zoo project. Their appearance is like that of its namesake, small elliptical and green in colour. This green colour is due to the strong optical emission lines of $\rm{[OIII]}$ and they display some of the highest specific star formation rates (sSFR) in the local Universe \citep{Cardamoneetal2009, Amorinetal2015}. The Green Beans are similar to Green Peas, being also sites of extremely high sSFR but they are larger and less compact than the Green Peas. Green Beans also have morphology that resembles that of Type 2 AGN, further supported by spectroscopic data indicating AGN like emission ratios resembling Type 2 AGN. 

\begin{figure}
\begin{centering}
	\includegraphics[width=\columnwidth]{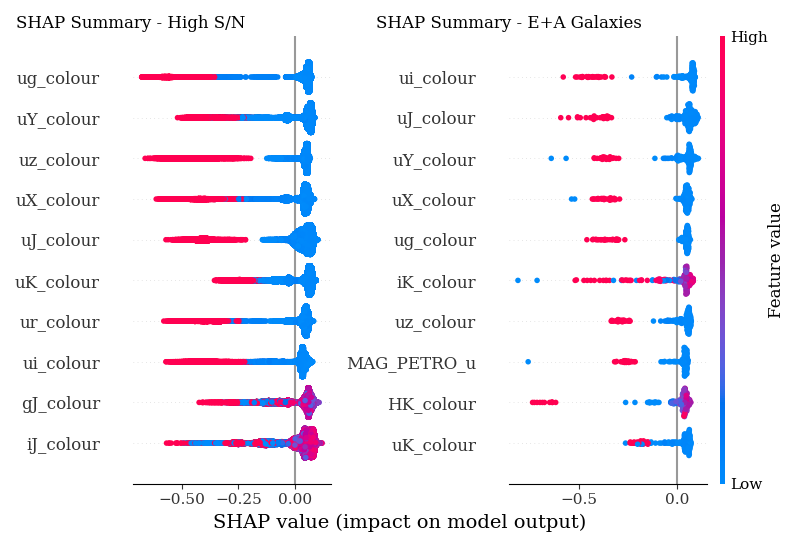}
    \caption{A summary plot created using the {\sc{shap}} Python tool. The plot shows the 10 most impactful features from the photometric data set in descending order. Left: High S/N sample. Right: E+A sample. We see in both samples that the u band colours have the greatest impact on the anomaly scores.}
    \label{fig:shap_photo}
\end{centering}
\end{figure}

We isolate 18 red spirals, most commonly isolated from the photometric isolation. These galaxies were initially observed by \citet{Mastersetal2010} and they are though to make up $\sim$6\% of all late-type spirals. Further study showed that red spirals are more common with stellar masses above $\rm{10^{10}M\odot}$ and become less common with decreasing stellar mass \citep{Maseaetal2014}. By evaluating the stellar masses of the red spirals we have isolated, we find that 16 have stellar masses $\rm{>10^{10}M\odot}$. This population of red spirals makes up approximately 35\% of isolated high S/N galaxies which is higher than we expected. Using the Python library {\sc{shap}}, we can visualise which features have the most impact when designating anomalies. From the photometric analysis shown in figure \ref{fig:shap_photo}, we can see that the u band colours and magnitudes are the most impactful features for designating anomalous objects. This tells us that the u band photometry is the most unique feature for this anomalous high S/N galaxies from the photometric isolation.

\subsection{Anomalous E+A Galaxies}
\label{sec:psbresults}

The most interesting outcome of the E+A isolations is the strong $\rm{H\alpha}$ galaxies. We find that 29 of the E+A anomalies have have strong $\rm{H\alpha}$ emission (EW$_{\rm{H\alpha}} >5$\text{\AA}), 20 from the spectroscopic isolation and 9 from the photometric isolation. This strong $\rm{H\alpha}$ is typically suggest of current star-formation \citep{Kennicutt1988a}, it could also be attributed to AGN activity \citep{Wilkinsonetal2017, Pawliketal2018, Greene2021} or could be a sign of dust obscuration \citep{Gotoetal2003, Pawliketal2018}.

AGNs can be identified by broad emission lines, most commonly $\rm{H\alpha}$ as well as $\rm{NII}$ \citep{BPT, Kewley2006}. By studying the spectra of these star-forming E+A galaxies, we initially find that 9 of the anomalies display $\rm{EW_{[OIII]} > 10}$\text{\AA} and as such, we broadly flag these galaxies as possible AGN hosts. Furthermore, we examine BPT and WHAN plots, shown in figure \ref{fig:bptwhan}, and we can see that there are only 2 possible AGN candidates. Checking the CATAIDs of these 2 objects, we see they are also in the same sample of strong $\rm{[OIII]}$. Looking again at the BPT-WHAN plot, we note that $\approx85$\% of the anomalous E+A galaxies lie within the star-forming delimiting region. Typically we would expect E+A galaxies to lie within the passive region \citep{Greene2021}. \citet{Greene2021} argues that the E+A galaxies found within this region are still undergoing inside out queching and as such, they are uncertain as to whether or not they can be classed as E+A galaxies. \citet{Pawliketal2018} however, argues that these strong $\rm{H\alpha}$ galaxies are a part of the evolutionary process of star-bursting galaxy to quenched elliptical and designates them as ePSBs. After analysing the BPT-WHAN plots and the spectra of our anomalous E+A galaxies, we conclude that 20 galaxies have strong $\rm{H\alpha}$ and do not exhibit possibilities of AGN hosts.

\citet{Gotoetal2003} studies the presence of strong $\rm{H\alpha}$ in E+A galaxies, compiling a sample of approximately 3000 E+A galaxies. These E+A galaxies have strong Balmer absorption and show little to no $\rm{[OII]}$. Of the 3000 galaxies, \citet{Gotoetal2003} find the 52\% have non-insignificant $\rm{H\alpha}$ emission and they suggest that these are dusty star-formers polluting the data set. \citet{Pawliketal2018} also delimits the previous ePSBs into a sub group of dPSBs, or dusty star-formers. These dPSBs are not, in-fact, post starburst galaxies but have been labelled such for convenience and are found by analysing the ratio between $\rm{H\alpha}$ and $\rm{H\beta}$. If $\rm{H\alpha / H\beta} > 2.87$ then \citet{Pawliketal2018} states they are dust obscured. We check our 20 E+A for dust obscuration and note that 7 of the objects are dust obscured according to the $\rm{H\alpha / H\beta}$ ratio, resulting in 13 E+A galaxies that are star-forming but not dust obscured. We want to further check for dust obscuration and analyse the colour-magnitude diagrams of the $g$ and $r$ bands. Using the definition by \citet{Shearman2014} note than none of the galaxies are anomalously red. Further study of the optical $g,r$ and infra-red $i$ bands agrees that there are no more dusty star-formers.

The spectra of these 13 star-forming E+A galaxies (SFE+A) suggests that they are not forming high-mass stars ($M\ge8M_\odot$) like O and B class, but are still efficiently forming lower mass stars ($M\le8M_\odot$, i.e. A,F,G,K \& M class stars). The lack of $\rm{[OII]}$ and $\rm{[OIII]}$ forbidden lines but the presence of strong $\rm{H\alpha}$ supports the notion that star-formation is happening, but massive, hot stars are not being produced.

\subsection{Small-scale Interactions}

The formation of low-mass stars is a well understood and observed process \citep{Larson1969, Shuetal1987, Swift2008} and they are believed to accrete much of their mass before nuclear burning can begin \citep{Shuetal1987}. High-mass stars however, have Kelvin-Helmholtz timescales that are less that their dynamical time, meaning the process of nuclear burning begins before all of their mass has been accreted \citep{McNally1964, BodenSweig, Larson1969}. This nuclear burning would cause an immense radiative pressure that repels the still accreting mass, suggesting a much faster accretion rate is needed for higher mass stars than lower mass stars \citep{YorkeSonn, KetoWood2006, Petersetal2010, Kuiperetal2011}. Observations of high-mass objects \citep{KetoWood2006}, as well as simulated scenarios \citep{Haemmerle2015}, shows that large accretion disks and rapid accretion rates can form high-mass stars in dense environments. We note that all of the identified SF E+A objects are low-mass and diffuse objects and suggest that recent small-scale interactions has caused a burst in star-formation that has not formed O or B class stars. We propose that this is due to the diffuse, gas poor galaxies not being able to accrete the necessary material for high mass stars before the radiative pressure expels the accreting material.

We analyse the morphology of our anomalous E+A galaxies using the parameters discussed in section \ref{sec:MorphParam}. In particular we use $G, M_{20}$ and $A$ to find merger signatures. Initial visual inspection showed that that all of the anomalous E+A galaxies are small, faint objects, with 80\% being in cluster environments. We use the $G-M_{20}$ plot shown in figure \ref{fig:GM20} to quantise each galaxy morphology. Only 8 anomalous E+A galaxies lie within the merger region of the plot, with a further 3 lying in the E/S0/Sa region and the remainder lying in the Sb/Sc/Irr region. We note that typical E+A galaxies should fall in the E/S0/Sa region, having elliptical morphologies. The 3 spectroscopically isolated E+A that lie in the merger region are all part of the SFE+A population, but remaining 10 SFE+As all fall in the Sb/Sc/Irr region. This suggests that perhaps they are not merging, but it does support that they are undergoing minor interactions causing the irregular morphology.

\subsection{Low Jeans Limits}

\citet{Steinhardt2024} proposed the idea that galaxies can still be star-forming, even though they appear by colour to be quiescent. This theory arose due to two issues in local large galaxies, namely; i) most observed massive blue star-forming galaxies that are presumed to be nearing the end of their stellar mass growth are $\sim$1dex less massive than local massive quiescent galaxies, and ii) the quiescent galaxies can not have gained this stellar mass from interactions or mergers since their mass ratios between stellar mass and black hole mass ($\rm{M_*/M_{BH}}$) are also $\sim$1dex less massive than expected \citep{Steinhardt2024}. M87 is taken as an example of a local massive quiescent galaxy. The stellar mass of M87 is $\rm{M_* = 10^{12.3}M_\odot}$ and the black hole mass is $\rm{M_{BH} = 10^{9.8}M_\odot}$. This stellar mass is $\sim$1dex larger than the largest known star-forming galaxies, suggesting that its stellar mass must have grown after its supposed turnoff. Furthermore, the $\rm{M_*/M_{BH}}$ ratio is much higher than theoretical work predicts \citep{KormendyHo2013}. \citet{Steinhardt2024} proposes then, that much of the final stellar mass is formed not during its blue phase but instead during its red phase, coining these galaxies, red star forming galaxies (RSFGs). Our proposed SFE+A galaxies could be a part of this population of ongoing star-forming galaxies as \citet{Steinhardt2024}'s proposed RSFGS would appear, spectroscopically, very similar to E+A galaxies, with strong $\rm{H\delta}$, low $\rm{[OII]}$ and non-negligible $\rm{H\alpha}$.

\citet{Steinhardt2024} provides a theory as to why these galaxies are not forming large mass stars, stating that the Jeans mass is so low that it is not possible for more massive stars to form. For a typical galactic molecular cloud, the speed of sound is $\sim$0.2 $\rm{km/s}$ \citep{KrumholzMckee}, and has a density of $\sim$100 $\rm{gcm^{-3}}$, giving a Jeans mass of $\rm{10M_\odot^2}$. As the temperature of the cloud drops and metallicity increases, the Jeans mass will also decrease. At a certain temperature and metallicity, the Jeans mass will be so low as to be unable to form O and B class stars. This would not completely prevent star formation as theoretically, AFGKM class stars could still efficiently form.

\subsection{Data Reduction Errors}

We also discuss the errors that the iForest algorithm has uncovered. All of these objects have been isolated from the spectroscopic datasets and we will discuss the reasons behind this shortly. We have identified 2 objects that are shredded galaxies, where a larger galaxy is falsely segmented into smaller galaxies. An example of the shredded galaxy is displayed in figure \ref{fig:Stripped} and it was identified by the iForest as it is an extremely bright star-forming sub-structure in a larger galaxy. The iForest found 13 `bad' spectra in the dataset, including 9 data reduction errors (example shown in figure \ref{fig:falseemission}) and 4 false emission line spectra (examples shown in figure \ref{fig:falseemission}). The iForest detects the sinusodial spectra due to errors in the EW measures which are flagged as unusual only in the spectroscopic data. The false emission lines do not effect the EW measures directly but they are instead detected as the far left side of the spectra dips below the continuum, causing errors in some of the EW measures. We study these false emission lines and find that they do not fall on any specific spectral lines. We analyse them further and find that all of these false emission line were observed on the same date (5th March 2008) but originate in different fibres, ruling out fibre issues. We plot the coordinates of each object and find no obvious correlation, making us rule out a sky phenomena such as a comet or other object. Though communications with S. Croom, we conclude that the errors are due to poor flat fielding and sky subtraction (Croom, priv.\ comm.). At the blue end of the spectra, flat fielding leads to zero divisions when normalising and could cause the line. Towards the red end of the spectra, flat fielding could be caused by fibre fringing which was an issue in early AAOmega that has since been fixed \citep{Sharpetal2006}.

\section{Conclusions}
We have utilised an iForest algorithm to isolate anomalous galaxies in two core samples, E+A galaxies, selected in a traditional manner, and high S/N galaxies. We supplied the unsupervised model with spectroscopic and photometric data for $\approx10^5$ objects and we find several rare phenomena and a number of odd morphologies.

As highlighted in \citetalias{BaronPoz2016} and their outlier detection model, the unsupervised iForest in the work also allows us to automatically identify outliers from the data, without having to choose explicit cuts. The only free parameter in the model is the outlier fraction, all other model parameters are learned automatically through the data and feature names.

From an input of 287 E+A galaxies, we isolate 49 unique anomalous galaxies, we chose to extract 10\% of the sample per dataset to produce a statistically important sample size. We find 25 E+A anomalies from the spectroscopic data and 24 anomalies from the photometric data. From an input of $\approx10^5$ S/N$>8$ galaxies, we isolate 52 unique anomalous galaxies, with 20 anomalies isolated from the spectroscopic data, 22 from photometric data and 10 from the combined spectroscopic/photometric data. Studying the 101 galaxies we find unusual emission lines, data reduction errors, EELGS, red spirals and possible star-forming E+A galaxies.

We chose to select our E+A galaxies in the same manner as \citep{DressGunn1983} – with strong Balmer absorption and low $\rm{[OII]}$. Through the spectroscopic isolation, we find that a significant portion of the E+A galaxies display extreme $\rm{H\alpha}$ emission. We initially theorised that this $\rm{H\alpha}$ emission could be caused by AGN activity. Visual inspection and a study of the $\rm{[OIII]}$ EW highlighted that of the 29 strong $\rm{H\alpha}$ E+As, 9 have strong $\rm{[OIII]}$ lines or show signs of an AGN. This result indicated that it was unlikely to be AGN activity leading to the strong $\rm{H\alpha}$. For further confirmation, we used BPT and WHAN \citep{BPT, WHAN} plots and find that there are very few AGN hosting E+A galaxies. We propose then, that these E+A galaxies are still star-forming, and this star-formation is what produces the $\rm{H\alpha}$ lines.

\citet{Gotoetal2003} suggests that these galaxies are dusty star-formers and the $\rm{[OII]}$ and $\rm{[OIII]}$ forbidden lines are obscured by dust. Using the definition by \citet{Shearman2014} we study the colour-mag ratios of the $g$ and $r$ bands and note than none of the galaxies are anomalously red. Study of the optical $g,r$ and infra-red $i$ bands suggests that these are not dusty star-formers so we use the method from \citet{Pawliketal2018}. \citet{Pawliketal2018} uses the ratio of $\rm{H\alpha/H\beta} > 2.87$ to confirm dust obscuration. We find that 7 of the E+A galaxies are dust obscured and are a separate population to the strong $\rm{[OIII]}$ population. This results in a total of 13 SFE+A galaxies that are not dusty, nor are they possible AGN hosts.

Strong $\rm{H\alpha}$ but no $\rm{[OII]}$ suggests that although star-formation is occurring, there are no high mass stars forming that can ionise the surrounding gas. If no high mass stars are forming, we provide two solutions: i) Small-scale interactions have perturbed the gas poor galaxies, resulting in some star-formation, but the radiative pressure of accreting massive stars prevents them from being formed, or ii) The E+A galaxies the iForest has identified are still star-forming efficiently but are unable to form O and B class stars due to higher metallicity and low temperatures.

This work shows that to select `pure' E+A galaxies, it is a necessity to select on $\rm{H\alpha}$ as well as $\rm{H\delta}$ and $\rm{[OII]}$. We can see, from the anomalies within the E+A selection method we chose, that the sample is polluted with dusty star-formers, possible AGN hosts and RSFGs. We also show the robustness of the iForest algorithm in detecting anomalies with no prior learning. We find over 50\% of the anomalies have not be identified in other works but we note that not all of the objects are scientifically interesting just because they are flagged as anomalies. This work could be expanded upon by utilising Active Anomaly Detection (AAD) to utilise expert input to tailor the models output. We can also test the hypotheses by studying the extended structure of the E+A galaxies to analyse how the galaxies might be quenching and interaction. We will probe this extended structure via image stacking and finding the averaged surface brightness profiles and the S\'ersic profile that best fits it.


\section*{Acknowledgements}

The anonymous reviewer is thanked for providing valuable feedback and revisions on the manuscript.

We thank C. Conselice for providing useful insights in to the analysis of our data.

We would also like to thank S. Croom for their assistance and correspondence that answered some of our unknowns.

KB would also wish to thank H. Green for assisting in classifying the visual morphology and for proof readings.

We would like to thank J. Liske who created all of the spectra shown in the paper figures, see in \citep{Liskeetal2015} for more detail. All of the PNG images were downloaded from \hyperlink{https://www.gama-survey.org/}{https://www.gama-survey.org/}.

This research made extensive use of {\sc{ipython}} \citep{PER-GRA:2007} and the following {\sc{python}} packages: astropy\footnote{\url{www.astropy.org/}}, a community-developed core Python package and an ecosystem of tools and resources for astronomy \citep{astropy:2013, astropy:2018, astropy:2022}; astroquery\footnote{\url{astroquery.readthedocs.io/en/latest/}}, a coordinated package for astropy \citep{Astroquery}; photutils\footnote{\url{https://photutils.readthedocs.io/en/stable/}}, a Python library that provides commonly-used tools and key functionality for detecting and performing photometry of astronomical sources \citep{photutils};
scikit-learn\footnote{\url{www.scikit-learn.org/}} \citep{Sklearn}; shap\footnote{{\url{https://shap.readthedocs.io/en/latest/}}}, a game theoretic approach to explain the output of any machine learning model \citep{shap}; and statmorph\footnote{\url{statmorph.readthedocs.io/en/latest/overview.html}}, an affiliated packed of astropy for calculating non-parametric morphological diagnostics of galaxy images (e.g., $G$-$M_{20}$ and $CAS$ statistics), as well as fitting 2D Sérsic profiles \citep{RodriquezGomez}.

\section*{Data Availability}

This work made extensive use of the GAMA DR4 data. GAMA is a joint European-Australasian project based around a spectroscopic campaign using the Anglo-Australian Telescope. The GAMA input catalogue is based on data taken from the Sloan Digital Sky Survey and the UKIRT Infrared Deep Sky Survey. Complementary imaging of the GAMA regions is being obtained by a number of independent survey programmes including GALEX MIS, VST KiDS, VISTA VIKING, WISE, Herschel-ATLAS, GMRT and ASKAP providing UV to radio coverage. GAMA is funded by the STFC (UK), the ARC (Australia), the AAO, and the participating institutions. The GAMA website is \hyperlink{https://www.gama-survey.org/}{https://www.gama-survey.org/}. We make specific use of {\sc{speclinessfr}} \citep{Gordonetal2017}, which compiles a significant amount of spectroscopic data and EW measures. We further use {\sc{gkvsciencecat}} \citep{Bellstedtetal2020}, which compiles the main survey selection including redshifts and photometry data. Lastly, we utilise {\sc{speccat}} and {\sc{apmatchedphotom}} \citep{Liskeetal2015} for extracting the spectra plots and for further photometry measures. We made extensive use of th GAMA Single Object Viewer\footnote{{\url{https://www.gama-survey.org/dr4/tools/sov.php}}}.



\bibliographystyle{mnras}
\bibliography{Reference} 

\begin{thebibliography}{}
\makeatletter
\relax
\def\mn@urlcharsother{\let\do\@makeother \do\$\do\&\do\#\do\^\do\_\do\%\do\~}
\def\mn@doi{\begingroup\mn@urlcharsother \@ifnextchar [ {\mn@doi@} {\mn@doi@[]}}
\def\mn@doi@[#1]#2{\def\@tempa{#1}\ifx\@tempa\@empty \href {http://dx.doi.org/#2} {doi:#2}\else \href {http://dx.doi.org/#2} {#1}\fi \endgroup}
\def\mn@eprint#1#2{\mn@eprint@#1:#2::\@nil}
\def\mn@eprint@arXiv#1{\href {http://arxiv.org/abs/#1} {{\tt arXiv:#1}}}
\def\mn@eprint@dblp#1{\href {http://dblp.uni-trier.de/rec/bibtex/#1.xml} {dblp:#1}}
\def\mn@eprint@#1:#2:#3:#4\@nil{\def\@tempa {#1}\def\@tempb {#2}\def\@tempc {#3}\ifx \@tempc \@empty \let \@tempc \@tempb \let \@tempb \@tempa \fi \ifx \@tempb \@empty \def\@tempb {arXiv}\fi \@ifundefined {mn@eprint@\@tempb}{\@tempb:\@tempc}{\expandafter \expandafter \csname mn@eprint@\@tempb\endcsname \expandafter{\@tempc}}}

\bibitem[\protect\citeauthoryear{Abraham, van~der Bergh  \& Nair}{Abraham et~al.}{2003}]{Abraham2003}
Abraham R.~G.,  van~der Bergh S.,   Nair P.,  2003, \mn@doi [ApJ] {10.1086/373919}, 588, 218

\bibitem[\protect\citeauthoryear{Adame et~al.,}{Adame et~al.}{2024}]{DESI}
Adame A.~G.,  et~al., 2024, \mn@doi [A&A] {10.3847/1538-3881/ad3217}, 168, 58

\bibitem[\protect\citeauthoryear{Amorín et~al.,}{Amorín et~al.}{2015}]{Amorinetal2015}
Amorín R.,  et~al., 2015, \mn@doi [A&A] {10.1051/0004-6361/201322786}, 578, A105

\bibitem[\protect\citeauthoryear{{Astropy Collaboration} et~al.,}{{Astropy Collaboration} et~al.}{2013}]{astropy:2013}
{Astropy Collaboration} et~al., 2013, \mn@doi [\aap] {10.1051/0004-6361/201322068}, \href {http://adsabs.harvard.edu/abs/2013A%26A...558A..33A} {558, A33}

\bibitem[\protect\citeauthoryear{{Astropy Collaboration} et~al.,}{{Astropy Collaboration} et~al.}{2018}]{astropy:2018}
{Astropy Collaboration} et~al., 2018, \mn@doi [\aj] {10.3847/1538-3881/aabc4f}, \href {https://ui.adsabs.harvard.edu/abs/2018AJ....156..123A} {156, 123}

\bibitem[\protect\citeauthoryear{{Astropy Collaboration} et~al.,}{{Astropy Collaboration} et~al.}{2022}]{astropy:2022}
{Astropy Collaboration} et~al., 2022, \mn@doi [\apj] {10.3847/1538-4357/ac7c74}, \href {https://ui.adsabs.harvard.edu/abs/2022ApJ...935..167A} {935, 167}

\bibitem[\protect\citeauthoryear{Baldwin, Phillips  \& Terlevich}{Baldwin et~al.}{1981}]{BPT}
Baldwin J.~A.,  Phillips M.~M.,   Terlevich R.,  1981, \mn@doi [PASP] {10.1086/130766}, 93, 5

\bibitem[\protect\citeauthoryear{Baron \& Poznanski}{Baron \& Poznanski}{2016}]{BaronPoz2016}
Baron D.,  Poznanski D.,  2016, \mn@doi [MNRAS] {10.1093/mnras/stw3021}, 465, 4530

\bibitem[\protect\citeauthoryear{{Bellstedt} et~al.,}{{Bellstedt} et~al.}{2020}]{Bellstedtetal2020}
{Bellstedt} S.,  et~al., 2020, \mn@doi [\mnras] {10.1093/mnras/staa1466}, 496, 3235

\bibitem[\protect\citeauthoryear{Bershady, Jangren  \& Conselice}{Bershady et~al.}{2000}]{Bershadyetal2000}
Bershady M.~A.,  Jangren A.,   Conselice C.~J.,  2000, \mn@doi [AJ] {10.1086/301386}, 119, 2645

\bibitem[\protect\citeauthoryear{Bodenheimer \& Sweigart}{Bodenheimer \& Sweigart}{1968}]{BodenSweig}
Bodenheimer P.,  Sweigart A.,  1968, \mn@doi [ApJ] {10.1086/149568}, 152, 515

\bibitem[\protect\citeauthoryear{Bradley et~al.,}{Bradley et~al.}{2024}]{photutils}
Bradley L.,  et~al., 2024, \mn@doi [Zenodo] {10.5281/zenodo.13989456}

\bibitem[\protect\citeauthoryear{{Cardamone} et~al.,}{{Cardamone} et~al.}{2009}]{Cardamoneetal2009}
{Cardamone} C.,  et~al., 2009, \mn@doi [MNRAS] {10.1111/j.1365-2966.2009.15383.x}, 399, 1191

\bibitem[\protect\citeauthoryear{Chang, Hsieh, Wang, Lin, Lim, Toba, Zhong  \& Chang}{Chang et~al.}{2021}]{Changetal2021}
Chang Y.-Y.,  Hsieh B.,  Wang W.-H.,  Lin Y.-T.,  Lim C.-F.,  Toba Y.,  Zhong Y.,   Chang S.-Y.,  2021, \mn@doi [ApJ] {10.3847/1538-4357/ac167c}, 920, 68

\bibitem[\protect\citeauthoryear{Chen et~al.,}{Chen et~al.}{2019}]{Chen2019}
Chen Y.-M.,  et~al., 2019, \mn@doi [MNRAS] {10.1093/mnras/stz2494}, 489, 5709

\bibitem[\protect\citeauthoryear{Cid~Fernandes, Stasinska, Schlickmann, Mateus, Vale~Asari, Schoenell  \& Sodre}{Cid~Fernandes et~al.}{2010}]{WHAN}
Cid~Fernandes r.,  Stasinska G.,  Schlickmann M.~S.,  Mateus A.,  Vale~Asari N.,  Schoenell W.,   Sodre L.,  2010, \mn@doi [MNRAS] {10.1111/j.1365-2966.2009.16185.x}, 403, 1036

\bibitem[\protect\citeauthoryear{Clarke, Scaife, Greenhalgh  \& Griguta}{Clarke et~al.}{2020}]{Clarketal2020}
Clarke A.~O.,  Scaife A. M.~M.,  Greenhalgh R.,   Griguta V.,  2020, \mn@doi [A&A] {10.1051/0004-6361/201936770}, 639, A84

\bibitem[\protect\citeauthoryear{Conselice}{Conselice}{2003}]{Conselice2003}
Conselice C.,  2003, \mn@doi [ApJS] {10.1086/375001}, 147, 1

\bibitem[\protect\citeauthoryear{Couch \& Sharples}{Couch \& Sharples}{1987}]{CouchSharp1987}
Couch W.~J.,  Sharples R.~M.,  1987, \mn@doi [MNRAS] {10.1093/mnras/229.3.423}, 229, 423

\bibitem[\protect\citeauthoryear{Dressler \& Gunn}{Dressler \& Gunn}{1982}]{DressGunn1982}
Dressler A.,  Gunn J.~E.,  1982, \mn@doi [ApJ] {10.1086/160524}, 263, 533

\bibitem[\protect\citeauthoryear{Dressler \& Gunn}{Dressler \& Gunn}{1983}]{DressGunn1983}
Dressler A.,  Gunn J.~E.,  1983, \mn@doi [ApJ] {10.1086/161093}, 270, 7

\bibitem[\protect\citeauthoryear{Driver et~al.,}{Driver et~al.}{2022}]{Driveretal2022}
Driver S.,  et~al., 2022, \mn@doi [MNRAS] {10.1093/mnras/stac472}, 513, 439

\bibitem[\protect\citeauthoryear{{Euclid Collaboration} et~al.,}{{Euclid Collaboration} et~al.}{2024}]{Euclid}
{Euclid Collaboration} et~al., 2024, \mn@doi [arXiv e-prints] {10.48550/arXiv.2405.13491}, p. arXiv:2405.13491

\bibitem[\protect\citeauthoryear{Freeman, Izbicki, Lee, Newman, Conselice, Koekemoer, Lotz  \& Mozena}{Freeman et~al.}{2013}]{Freeman2013}
Freeman P.~E.,  Izbicki R.,  Lee A.~B.,  Newman J.~A.,  Conselice C.~J.,  Koekemoer A.~M.,  Lotz J.~M.,   Mozena M.,  2013, \mn@doi [MNRAS] {10.1093/mnras/stt1016}, 434, 282

\bibitem[\protect\citeauthoryear{{Ginsburg} et~al.,}{{Ginsburg} et~al.}{2019}]{Astroquery}
{Ginsburg} A.,  et~al., 2019, \mn@doi [\aj] {10.3847/1538-3881/aafc33}, 157, 98

\bibitem[\protect\citeauthoryear{{Goel} \& {Montgomery}}{{Goel} \& {Montgomery}}{2015}]{AmitMichele2015}
{Goel} A.,  {Montgomery} M.,  2015, in IAU General Assembly. p. 2255500

\bibitem[\protect\citeauthoryear{Gordon et~al.,}{Gordon et~al.}{2017}]{Gordonetal2017}
Gordon Y.,  et~al., 2017, \mn@doi [MNRAS] {10.1093/mnras/stw2925}, 465, 2671

\bibitem[\protect\citeauthoryear{Goto}{Goto}{2007}]{Goto2007a}
Goto T.,  2007, \mn@doi [MNRAS] {10.1111/j.1365-2966.2007.11674.x}, 377, 1222

\bibitem[\protect\citeauthoryear{Goto et~al.,}{Goto et~al.}{2003}]{Gotoetal2003}
Goto T.,  et~al., 2003, \mn@doi [PASJ] {10.1093/pasj/55.4.771}, 55, 771

\bibitem[\protect\citeauthoryear{Greene, Anderson, Marinelli, Holley-Bockelmann, Campbel  \& Liu}{Greene et~al.}{2021}]{Greene2021}
Greene O.~A.,  Anderson M.~R.,  Marinelli M.,  Holley-Bockelmann K.,  Campbel L. E.~P.,   Liu C.~T.,  2021, \mn@doi [ApJ] {10.3847/1538-4357/ae4d1}, 910, 162

\bibitem[\protect\citeauthoryear{Haemmerle, Eggenberger, Meynet, Maeder  \& Charbonnel}{Haemmerle et~al.}{2015}]{Haemmerle2015}
Haemmerle L.,  Eggenberger P.,  Meynet G.,  Maeder A.,   Charbonnel C.,  2015, \mn@doi [A&A] {10.1051/0004-6361/201527202}, 585, A65

\bibitem[\protect\citeauthoryear{Hogg, Masjedi, Berlind, R, Quintero  \& J}{Hogg et~al.}{2006}]{Hoggetal2006}
Hogg D.~W.,  Masjedi M.,  Berlind A.~A.,  R B.~M.,  Quintero A.~D.,   J B.,  2006, \mn@doi [ApJ] {10.1086/507172}, 650, 763

\bibitem[\protect\citeauthoryear{{Holwerda} et~al.,}{{Holwerda} et~al.}{2025}]{Holwerda2025}
{Holwerda} B.~W.,  et~al., 2025, \mn@doi [PASA] {10.1017/pasa.2025.5}, 42, e028

\bibitem[\protect\citeauthoryear{Hopkins, Hernquist, Cox  \& Kere{\v{s}}}{Hopkins et~al.}{2008}]{Hopkinsetal2008}
Hopkins P.~F.,  Hernquist L.,  Cox T.~J.,   Kere{\v{s}} D.,  2008, \mn@doi [ApJS] {10.1086/524362}, 175, 356

\bibitem[\protect\citeauthoryear{{Iglesias-Páramo, J.} et~al.,}{{Iglesias-Páramo, J.} et~al.}{2022}]{Iglesias2022}
{Iglesias-Páramo, J.} et~al., 2022, \mn@doi [A&A] {10.1051/0004-6361/202243931}, 665, A95

\bibitem[\protect\citeauthoryear{Ishida et~al.,}{Ishida et~al.}{2021}]{Ishida2021}
Ishida E. E.~O.,  et~al., 2021, \mn@doi [A&A] {10.1051/0004-6361/202037709}, 650, A195

\bibitem[\protect\citeauthoryear{{Ivezi{\'c}} et~al.,}{{Ivezi{\'c}} et~al.}{2019}]{LSSTRubin}
{Ivezi{\'c}} {\v{Z}}.,  et~al., 2019, \mn@doi [ApJ] {10.3847/1538-4357/ab042c}, 873, 111

\bibitem[\protect\citeauthoryear{Kamalov \& Leung}{Kamalov \& Leung}{2020}]{Kamalov2019}
Kamalov F.,  Leung H.~H.,  2020, \mn@doi [JIKM] {10.1142/S0219649220400134}, 19, 2040013

\bibitem[\protect\citeauthoryear{{Kennicutt}}{{Kennicutt}}{1998}]{Kennicutt1988a}
{Kennicutt} Jr. R.~C.,  1998, \mn@doi [\araa] {10.1146/annurev.astro.36.1.189}, 36, 189

\bibitem[\protect\citeauthoryear{Keto \& Wood}{Keto \& Wood}{2006}]{KetoWood2006}
Keto E.,  Wood K.,  2006, \mn@doi [ApJ] {10.1086/498611}, 637, 850

\bibitem[\protect\citeauthoryear{Kewley, Groves, Kauffmann  \& Heckman}{Kewley et~al.}{2006}]{Kewley2006}
Kewley L.~J.,  Groves B.,  Kauffmann G.,   Heckman T.,  2006, \mn@doi [MNRAS] {10.1111/j.1365-2966.2006.10859.x}, 372, 961

\bibitem[\protect\citeauthoryear{{Kormendy} \& {Ho}}{{Kormendy} \& {Ho}}{2013}]{KormendyHo2013}
{Kormendy} J.,  {Ho} L.~C.,  2013, \mn@doi [\araa] {10.1146/annurev-astro-082708-101811}, 51, 511

\bibitem[\protect\citeauthoryear{Krumholz \& McKee}{Krumholz \& McKee}{2008}]{KrumholzMckee}
Krumholz M.~R.,  McKee C.~F.,  2008, \mn@doi [Nature] {10.1038/nature06620}, 451, 1082

\bibitem[\protect\citeauthoryear{Kuiper, Klahr, Beuther  \& Henning}{Kuiper et~al.}{2011}]{Kuiperetal2011}
Kuiper R.,  Klahr H.,  Beuther H.,   Henning T.,  2011, \mn@doi [ApJ] {10.1088/004-637X/732/1/20}, 732, 11

\bibitem[\protect\citeauthoryear{Larson}{Larson}{1969}]{Larson1969}
Larson R.~B.,  1969, \mn@doi [MNRAS] {10.1093/mnras/145.3.271}, 145, 271

\bibitem[\protect\citeauthoryear{Liske et~al.,}{Liske et~al.}{2015}]{Liskeetal2015}
Liske J.,  et~al., 2015, \mn@doi [MNRAS] {10.1093/mnras/stv1436}, 452, 2087

\bibitem[\protect\citeauthoryear{Liu, Ting  \& Zhou}{Liu et~al.}{2008}]{IsolationForest}
Liu F.~T.,  Ting K.~M.,   Zhou Z.-H.,  2008, in 2008 Eighth IEE International Conference on Data Mining. pp 413--422, \mn@doi{10.1109/ICDM.2008.17}

\bibitem[\protect\citeauthoryear{Liu, Li, Li  \& Zhang}{Liu et~al.}{2018}]{Liuetal2018}
Liu H.,  Li X.,  Li J.,   Zhang S.,  2018, \mn@doi [IEEE Trans. Syst. Man. Cybern.] {10.1109/TSMC.2017.2718220}, 48, 2451

\bibitem[\protect\citeauthoryear{Llerna et~al.,}{Llerna et~al.}{2024}]{Llerenaetal2024}
Llerna M.,  et~al., 2024, \mn@doi [A&A] {10.1051/0004-6361/202449904}, 691, A59

\bibitem[\protect\citeauthoryear{Lochner, McEwen, Peiris, Lahav  \& Winter}{Lochner et~al.}{2016}]{Lochneretal2016}
Lochner M.,  McEwen J.~D.,  Peiris H.~V.,  Lahav O.,   Winter M.~K.,  2016, \mn@doi [ApJS] {10.3847/0067-0049/225/2/31}, 225, 31

\bibitem[\protect\citeauthoryear{Lotz, Primack  \& Madau}{Lotz et~al.}{2004}]{Lotzetal2004}
Lotz J.,  Primack J.,   Madau P.,  2004, \mn@doi [AJ] {10.1086/421849}, 128, 163

\bibitem[\protect\citeauthoryear{{Lotz} et~al.,}{{Lotz} et~al.}{2008}]{Lotzetal2008}
{Lotz} J.~M.,  et~al., 2008, \mn@doi [ApJ] {10.1086/523659}, 672, 177

\bibitem[\protect\citeauthoryear{{Lumbreras-Calle} et~al.,}{{Lumbreras-Calle} et~al.}{2022}]{LumbrerasCalle2021}
{Lumbreras-Calle} A.,  et~al., 2022, \mn@doi [A&A] {10.1051/0004-6361/202142898}, 668, A60

\bibitem[\protect\citeauthoryear{Lundberg et~al.,}{Lundberg et~al.}{2020}]{shap}
Lundberg S.~M.,  et~al., 2020, Nature Machine Intelligence, 2, 2522

\bibitem[\protect\citeauthoryear{Margalef-Bentabol, Huertas-Compant, Charnock, Margalef-Bentabol, Bernardi, Dubois, Storey-Fisher  \& Zanisi}{Margalef-Bentabol et~al.}{2020}]{Margalefetal2020}
Margalef-Bentabol B.,  Huertas-Compant M.,  Charnock T.,  Margalef-Bentabol C.,  Bernardi M.,  Dubois Y.,  Storey-Fisher K.,   Zanisi L.,  2020, \mn@doi [MNRAS] {10.1093/mnras/staa1647}, 496, 2346

\bibitem[\protect\citeauthoryear{Maseda et~al.,}{Maseda et~al.}{2014}]{Maseaetal2014}
Maseda M.~C.,  et~al., 2014, \mn@doi [ApJ] {10.1088/0004-637X/791/1/17}, 791, 17

\bibitem[\protect\citeauthoryear{Masters et~al.,}{Masters et~al.}{2010}]{Mastersetal2010}
Masters K.~L.,  et~al., 2010, \mn@doi [MNRAS] {10.1111/j.1365-2966.2010.16503.x}, 405, 783

\bibitem[\protect\citeauthoryear{McNally}{McNally}{1964}]{McNally1964}
McNally D.,  1964, \mn@doi [ApJ] {10.1086/148007}, 140, 1088

\bibitem[\protect\citeauthoryear{Pawlik et~al.,}{Pawlik et~al.}{2018}]{Pawliketal2018}
Pawlik M.~M.,  et~al., 2018, \mn@doi [MNRAS] {10.1093/mnras/sty589}, 477, 1708

\bibitem[\protect\citeauthoryear{Pedregosa et~al.,}{Pedregosa et~al.}{2012}]{Sklearn}
Pedregosa F.,  et~al., 2012, J. Mach. Learn. Res., 12, 2825

\bibitem[\protect\citeauthoryear{P\'erez \& Granger}{P\'erez \& Granger}{2007}]{PER-GRA:2007}
P\'erez F.,  Granger B.~E.,  2007, \mn@doi [Computing in Science and Engineering] {10.1109/MCSE.2007.53}, 9, 21

\bibitem[\protect\citeauthoryear{Peters, Banerjee, Klessen, Mac~Low  \& Galvan-Madrid}{Peters et~al.}{2010}]{Petersetal2010}
Peters T.,  Banerjee R.,  Klessen R.~S.,  Mac~Low M.~M.,   Galvan-Madrid R~nd~Keto E.~R.,  2010, \mn@doi [ApJ] {10.1088/0004-637X/711/2/1017}, 711, 1017

\bibitem[\protect\citeauthoryear{Peth et~al.,}{Peth et~al.}{2015}]{Peth2015}
Peth M.~A.,  et~al., 2015, \mn@doi [MNRAS] {10.1093/mnras/stw252}, 458, 963

\bibitem[\protect\citeauthoryear{Poggianti et~al.,}{Poggianti et~al.}{2009}]{Poggiantietal2009}
Poggianti B.~M.,  et~al., 2009, \mn@doi [ApJ] {10.1088/0004-637X/693/1/112}, 693, 112

\bibitem[\protect\citeauthoryear{Prescott \& Sanderson}{Prescott \& Sanderson}{2019}]{PresSand2019}
Prescott M. K.~M.,  Sanderson K.~N.,  2019, \mn@doi [ApJ] {10.3847/1538-4357/ab44b1}, 885, 40

\bibitem[\protect\citeauthoryear{Reza}{Reza}{2021}]{Reza2021}
Reza M.,  2021, \mn@doi [Astron. Comput.] {10.1016/j.ascom.2021.100492}, 37, 100492

\bibitem[\protect\citeauthoryear{Rodriquez-Gomez et~al.,}{Rodriquez-Gomez et~al.}{2019}]{RodriquezGomez}
Rodriquez-Gomez V.,  et~al., 2019, \mn@doi [MNRAS] {10.1093/mnras/sty3345}, 483, 4140

\bibitem[\protect\citeauthoryear{Sharp et~al.,}{Sharp et~al.}{2006}]{Sharpetal2006}
Sharp R.,  et~al., 2006, \mn@doi [SPIE] {10.1117/12.671022}, 6269E, 14

\bibitem[\protect\citeauthoryear{Shearman \& Pimbblet}{Shearman \& Pimbblet}{2014}]{Shearman2014}
Shearman O.,  Pimbblet K.~A.,  2014, \mn@doi [PASA] {doi:10.1017/pasa.2014.34}, 31, 38

\bibitem[\protect\citeauthoryear{{Shu}, {Adams}  \& {Lizano}}{{Shu} et~al.}{1987}]{Shuetal1987}
{Shu} F.~H.,  {Adams} F.~C.,   {Lizano} S.,  1987, \mn@doi [\araa] {10.1146/annurev.aa.25.090187.000323}, 25, 23

\bibitem[\protect\citeauthoryear{Smit R~nd~Bouwens et~al.,}{Smit et~al.}{2015}]{Smitetal2015}
Smit R~nd~Bouwens R.~J.,  et~al., 2015, \mn@doi [ApJ] {10.1088/0004-637X/801/2/122}, 801, 122

\bibitem[\protect\citeauthoryear{{Steinhardt}}{{Steinhardt}}{2024}]{Steinhardt2024}
{Steinhardt} C.~L.,  2024, \mn@doi [arXiv e-prints] {10.48550/arXiv.2402.03423}, p. arXiv:2402.03423

\bibitem[\protect\citeauthoryear{{Swift} \& {Welch}}{{Swift} \& {Welch}}{2008}]{Swift2008}
{Swift} J.~J.,  {Welch} W.~J.,  2008, \mn@doi [\apjs] {10.1086/520846}, 174, 202

\bibitem[\protect\citeauthoryear{Tahi \& Hadi}{Tahi \& Hadi}{2019}]{TahiHadi2019}
Tahi A.,  Hadi A.~S.,  2019, \mn@doi [ACM Comput. Surv.] {10.1145/3312739}, 52, 1

\bibitem[\protect\citeauthoryear{Vergani et~al.,}{Vergani et~al.}{2010}]{Verganietal2010}
Vergani D.,  et~al., 2010, \mn@doi [A&A] {10.1051/0004-6361/200912802}, 509, A42

\bibitem[\protect\citeauthoryear{Wilkinson, Pimbblet  \& Stott}{Wilkinson et~al.}{2017}]{Wilkinsonetal2017}
Wilkinson C.,  Pimbblet K.,   Stott J.,  2017, \mn@doi [MNRAS] {10.1093/mnras/stx2034}, 472, 1447

\bibitem[\protect\citeauthoryear{{York} et~al.,}{{York} et~al.}{2000}]{Yorketal2000}
{York} D.~G.,  et~al., 2000, \mn@doi [\aj] {10.1086/301513}, 120, 1579

\bibitem[\protect\citeauthoryear{Yorke \& Sonnhalter}{Yorke \& Sonnhalter}{2002}]{YorkeSonn}
Yorke H.~W.,  Sonnhalter C.,  2002, \mn@doi [ApJ] {10.1086/339264}, 569, 846

\bibitem[\protect\citeauthoryear{Zabludoff, Zaritsky, Lin, Tucker, Hasimoto, Shectman, Oemler  \& Kirshner}{Zabludoff et~al.}{1996}]{Zabludoffetal1996}
Zabludoff A.~I.,  Zaritsky D.,  Lin H.,  Tucker D.,  Hasimoto Y.,  Shectman S.~A.,  Oemler A.,   Kirshner R.~P.,  1996, \mn@doi [ApJ] {10.1086/177495}, 466, 104

\makeatother
\end{thebibliography}



\appendix
\section{Summary}

In this appendix we summarise all of the outlying galaxies in tables displaying their coordinates, classification and the sample they have been isolated from and the data used in said isolation. The galaxy classification has been allocated through visual inspection by KB, while also taking into account the designation in the SDSS navigator, and a secondary inspection by H. Green was performed to confirm the classification.

\begin{figure*}
    \centering
	\includegraphics[width=\textwidth]{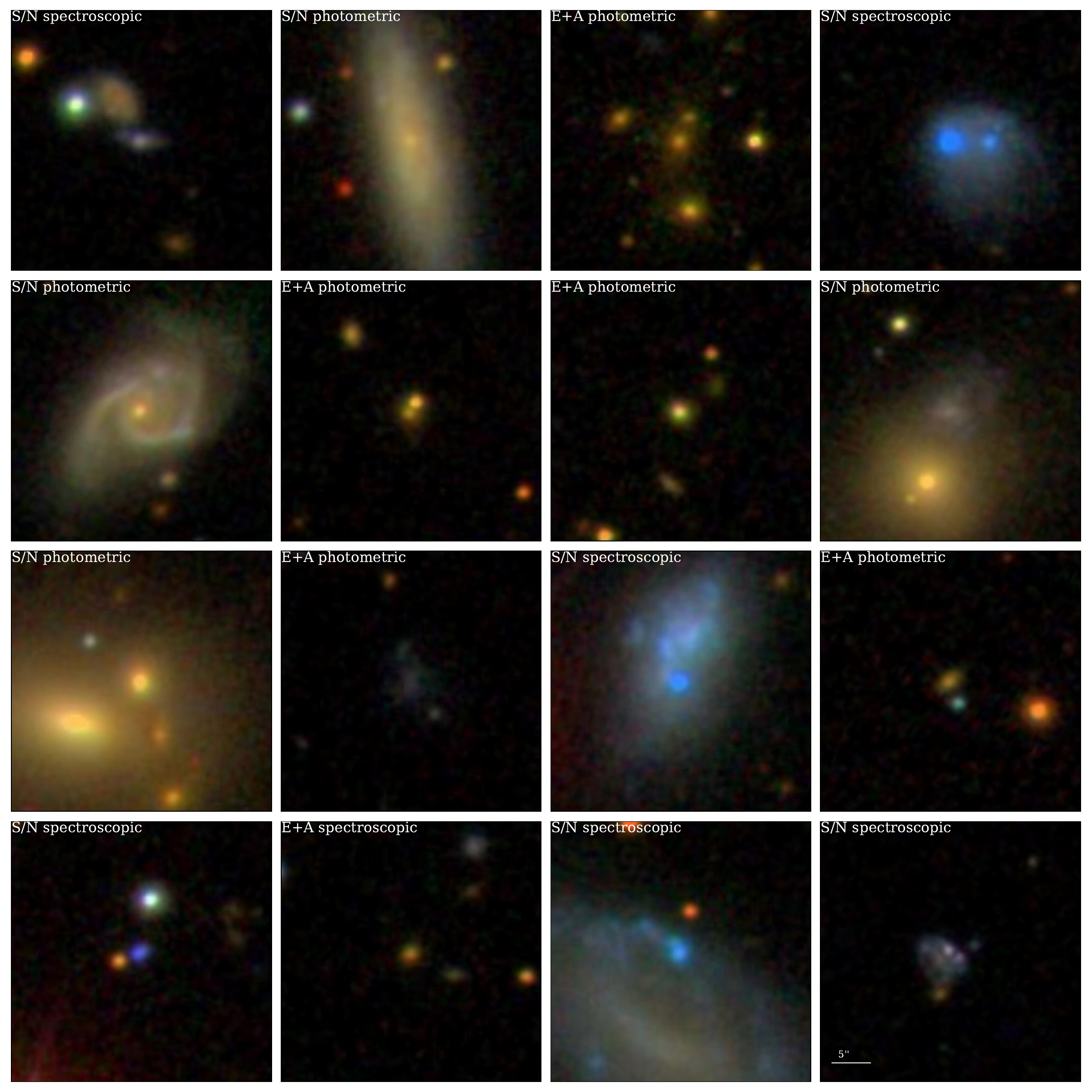}
    \caption{Shown above are the most anomalous 16 galaxies found from the CAS selection. Anomalies in this case are defined by plotting the CAS parameters and then finding the point density of the nearest 10 neighbours of the iForest galaxies.
    Shown in the top corner of each image is the sample the galaxy has been selected from. E+A and S/N shows the data selecting process and spectroscopic, photometric or combination shows the data provided to the iForest.}
    \label{fig:CASweird}
\end{figure*}

\begin{table*}
    \centering
    \caption{The anomalous objects found from the E+A isolations are highlighted in the table. We display the object name from GAMA where possible and where not possible an alternate name. We further display the RA and DEC coordinates in deg. The class column indicates the type of object, S = Spiral, S0, I = Irregular, M = Multiple Nuclei, E = Elliptical. The isolation data column indicates from which isolation the object has been selected from, S = Spectroscopic, P = Photometric and C = Combination. Where duplicate objects are found, we indicate all samples the object is isolated from.}
    \label{tab:EAtable}
    \begin{tabular}{lcccr}
    \hline
    \textbf{Galaxy   Name}       & \textbf{RA (deg)} & \textbf{DEC (deg)} & \textbf{Class} & \textbf{Isolation Data} \\
    \hline
        GAMA 623866 & 140.63771 & 0.81636 & S0 & S + P + C \\ 
        DESI J216.5852+01.6198 & 216.58525 & 1.61986 & S & S + C \\ 
        GAMA 250176 & 214.18704 & 1.97408 & I & S + C \\ 
        GAMA 319942 & 216.30633 & 1.84812 & S0 & S + C \\ 
        GAMA 599349 & 131.04029 & 0.22654 & S0 & S + C \\ 
        GAMA 098625 & 179.03608 & 0.9419 & S0 & S + C \\ 
        SDSS J142046.10+004641.8 & 215.19208 & 0.77829 & M & S + C \\ 
         SDSS J120756.28-020301.3 & 181.98454 & -2.05038 & I & S \\ 
         SDSS J120825.77-020134.1 & 182.10737 & -2.02615 & S0 & S \\ 
        GAMA 136501 & 175.18121 & -1.78885 & I & S \\ 
        GAMA 185670 & 181.19825 & -1.5217 & I & S \\ 
        GAMA 272241 & 178.43688 & 1.43316 & S0 & S \\ 
        GAMA 298897 & 222.12246 & 1.19695 & E & S \\ 
        GAMA 319949 & 216.36367 & 1.83371 & M & S \\ 
        GAMA 512767 & 219.90375 & -1.06983 & S0 & S \\ 
        GAMA 535853 & 179.51933 & -0.85615 & I & S \\ 
        GAMA 536399 & 182.154007 & -0.9574 & M & S \\ 
        GAMA 560613 & 180.7675 & -0.59587 & S0 & S \\ 
        GAMA 084615 & 178.84121 & 0.47832 & I & S \\ 
        GAMA 092370 & 216.0315 & 0.53364 & S & S \\ 
        GAMA 093959 & 223.16904 & 0.56275 & I & S \\ 
        SDSS J090230.75+003137.8 & 135.62817 & 0.5272 & M & S \\ 
        SDSS J090248.47-001215.5 & 135.70196 & -0.20432 & S & S \\ 
        SDSS J141533.72-005913.5 & 213.89054 & -0.9871 & E & S \\ 
        SDSS J141546.02-010922.1 & 213.94179 & -1.15615 & S0 & S \\ 
        GAMA 130823 & 178.82567 & -2.19445 & S0 & P + C \\ 
        GAMA 203840 & 135.44063 & -0.29678 & M & P + C \\ 
        GAMA 215981 & 136.03192 & 0.52556 & E & P + C \\ 
        GAMA 302687 & 138.82412 & 1.43235 & M & P + C \\ 
        GAMA 302729 & 139.07225 & 1.47696 & E & P + C \\ 
        GAMA 324714 & 136.92746 & 1.72548 & S & P + C \\ 
        GAMA 382668 & 137.91567 & 1.94651 & S0 & P + C \\ 
        GAMA 387344 & 135.94775 & 2.33613 & E & P + C \\ 
        GAMA 049759 & 222.42671 & -0.70777 & S0 & P + C \\ 
        GAMA 056197 & 184.84033 & -0.25884 & E & P + C \\ 
        GAMA 007568 & 178.24217 & 0.77447 & E & P + C \\ 
        GAMA 007758 & 179.35896 & 0.66938 & E & P + C \\ 
        SDSS J091504.74+024105.1 & 138.64029 & 1.73369 & M & P + C \\ 
        SDSS J092233.04+004858.8 & 133.21933 & -1.61439 & E & P + C \\ 
        SDSS J115518.16-021140.0 & 177.07012 & 0.64766 & E & P + C \\ 
        SDSS J141124.12+024258.4 & 212.8505 & 2.71623 & I & P + C \\ 
        WHL J085252.6-013652 & 139.83333 & -1.93921 & S0 & P + C \\ 
        WISEA J091919.99-015621.2 & 138.76975 & 2.68476 & E & P + C \\ 
        GAMA 208796 & 130.18838 & 0.03409 & S & P \\ 
        GAMA 278848 & 133.949997 & 0.81399 & E & P \\ 
        GAMA 302685 & 138.82104 & 1.37455 & M & P \\ 
        GAMA 550260 & 134.68558 & -0.44041 & E & P \\ 
        GAMA 601598 & 140.64658 & 0.36883 & S & P \\ 
        GAMA 694909 & 184.97717 & 0.58512 & S0 & P \\ 
        \hline
\end{tabular}
\end{table*}

\begin{table*}
    \centering
    \caption{The anomalous objects found from the S/N isolations are highlighted in the table. We display the object name from GAMA where possible and where not possible an alternate name. We further display the RA and DEC coordinates in deg. The class column indicates the type of object, S = Spiral, S0, I = Irregular, M = Multiple Nuclei, E = Elliptical and F = False flagged objects (not galaxies). The isolation data column indicates from which isolation the object has been selected from, S = Spectroscopic, P = Photometric and C = Combination. Where duplicate objects are found, we indicate all samples the object is isolated from.}
    \label{tab:SNtable}
\begin{tabular}{lcccr}
\hline
\textbf{Galaxy   Name}       & \textbf{RA (deg)} & \textbf{DEC (deg)} & \textbf{Class} & \textbf{Isolation Data} \\
\hline
        DESI J212.9710-01.9480 & 212.97275 & -0.05066 & S & C \\ 
        GAMA 345754 & 131.11858 & 2.06396 & I & C \\ 
        GAMA 521949 & 131.96658 & 2.92294 & M & C \\ 
        SDSS J114838.04-014557.9 & 177.15942 & -0.23458 & F & C \\ 
        SDSS J141504.11+021706.9 & 213.76712 & 2.28525 & S & C \\ 
        SDSS J144612.77+021834.7 & 221.55358 & 2.31047 & S & C \\ 
        WISEA J084735.34-010236.2 & 131.90192 & -0.95406 & S & C \\ 
        WISEA J085902.01-021322.3 & 134.75604 & -1.77504 & S & C \\ 
        WISEA J091248.93+024451.4 & 138.20425 & 1.97904 & S & C \\ 
        WISEA J092026.68-025859.9 & 140.11092 & -1.01865 & S0 & C \\ 
         SDSS J142439.90+024120.4 & 216.16625 & 2.68903 & S & P \\ 
        GAMA 345447 & 129.48546 & 2.02111 & S & P \\ 
        GAMA 365248 & 129.02217 & 1.89409 & S & P \\ 
        GAMA 375560 & 129.87421 & 1.23344 & E & P \\ 
        GAMA 380578 & 129.24721 & 1.79204 & S0 & P \\ 
        PGC1 0051957 GROUP & 218.11875 & 0.29401 & S & P \\ 
        SDSS J114006.78-010944.4 & 175.02667 & -0.83774 & E & P \\ 
        SDSS J140929.04-020544.0 & 212.37017 & -1.9064 & S & P \\ 
        WISEA J113902.88-014736.5 & 174.76383 & -0.206 & S & P \\ 
        GALEXASC J090337.86-025224.8 & 135.90387 & -1.13016 & I & P + C \\ 
        GAMA 085416 & 182.36933 & 0.53327 & S & P + C \\ 
        GAMA 214184 & 129.00642 & 0.43808 & S & P + C \\ 
        GAMA 278390 & 130.94883 & 0.83844 & S & P + C \\ 
        GAMA 372571 & 136.46367 & 1.13359 & S0 & P + C \\ 
        GAMA 517279 & 131.68037 & 2.53922 & M & P + C \\ 
        GAMA J141735.55+020311.9 & 214.39804 & 2.05322 & M & P + C \\ 
        SDSS-C4 1366 & 223.39608 & 0.01039 & S & P + C \\ 
        WISEA J090121.80-020859.2 & 135.34375 & -1.85325 & S & P + C \\ 
        WISEA J090330.87-025354.8 & 135.87854 & -1.10327 & S & P + C \\ 
        WISEA J090400.77-015455.3 & 136.00417 & -0.08876 & S & P + C \\ 
        WISEA J120843.40-012715.9 & 182.18421 & -0.54084 & E & P + C \\ 
        WISEA J143104.87-010449.7 & 217.76737 & -0.9191 & S & P + C \\ 
         SDSS J115133.34-022221.9 & 177.88896 & -2.37276 & M & S \\ 
        DESI J217.4525-01.1694 & 217.45254 & -1.16935 & M & S \\ 
        GAMA 006821 & 174.15312 & 0.81543 & M & S \\ 
        GAMA 136473 & 175.11621 & -1.63778 & I & S \\ 
        GAMA 137625 & 178.93579 & -1.79443 & E & S \\ 
        GAMA 210224 & 137.08458 & 0.19551 & I & S \\ 
        GAMA 376183 & 132.31346 & 1.48846 & S0 & S \\ 
        GAMA 418979 & 138.35871 & 2.74863 & I & S \\ 
        GAMA 691332 & 184.44154 & -0.78311 & E & S \\ 
        SDSS J084152.37+025336.2 & 130.46825 & 2.8934 & M & S \\ 
        SDSS J085218.91-010458.8 & 133.07883 & -1.08304 & S0 & S \\ 
        SDSS J115036.14-003410.2 & 177.65058 & -0.56951 & M & S \\ 
        SDSS J121713.09+013747.4 & 184.30454 & 1.62984 & S0 & S \\ 
        SDSS J141440.82-003826.5 & 213.67008 & -0.64071 & S0 & S \\ 
        SDSS J142223.42-002225.3 & 215.59762 & -0.37369 & F & S \\ 
        SDSS J142435.47-014638.2 & 216.14779 & -1.7773 & E & S \\ 
        SDSS J144148.48+004128.1 & 220.45208 & 0.69113 & F & S \\ 
        SDSS J145107.11+023125.0 & 222.77963 & 2.52362 & S0 & S \\ 
        WISEA J085226.85-010241.9 & 133.110992 & -1.04479 & M & S \\ 
        WISEA J085229.19-011710.0 & 133.12171 & -1.28604 & E & S \\ 
        WISEA J090937.59-010912.2 & 137.40617 & -1.15371 & M & S \\ 
        WISEA J091004.95-012910.3 & 137.52046 & -1.4863 & S0 & S \\ 
        \hline
\end{tabular}
\end{table*}

\bsp	
\label{lastpage}
\end{document}